\documentclass{aa}
\usepackage{epsfig,graphicx,lscape}
\usepackage{float,longtable,comment}
\usepackage{rotating,amsmath}
\usepackage{natbib}
\newcommand{\kms}{{{km\,s}$^{-1}$\,}}
\newcommand{\teff}{{$T_\mathrm{eff}$\,}}
\newcommand{\logg}{{log~$g$\,}}
\newcommand{\vsini}{$v\sin i\,$}
\newcommand{\Msun}{\,$\mathrm{M}_\odot$}

\begin{document}
\title{The VLT-FLAMES Tarantula Survey\\
V.  The peculiar B[e]-like supergiant, VFTS698, in 30~Doradus\thanks{Based on observations at the European Southern Observatory Very Large Telescope in programme 182.D-0222}}

\author{P.~R.~Dunstall\inst{1}, M.~Fraser\inst{1}, J.~S.~Clark\inst{2}, P.~A.~Crowther\inst{3}, P.~L.~Dufton\inst{1},  C.~J.~Evans\inst{4}, \\
D.~J.~Lennon\inst{5}, I.~Soszy\'nski\inst{6}, W.~D.~Taylor\inst{7}, J.~S.~Vink\inst{8}
}

\institute{Department of Physics \& Astronomy, The Queen's University 
of Belfast, BT7 1NN, Northern Ireland, UK
	\and Department of Physics and Astronomy, The Open University, Walton Hall, Milton Keynes, MK7 6AA, UK
	\and Department of Physics \& Astronomy, University of Sheffield, Sheffield S3 7RH, UK
	\and UK Astronomy Technology Centre, Royal Observatory Edinburgh, Blackford Hill, Edinburgh, EH9 3HJ, UK
	\and ESA, Space Telescope Science Institute, 3700 San Martin Drive, Baltimore, MD 21218, USA
	\and Warsaw University Observatory, Al. Ujazdowskie 4, 00-478 Warszawa, Poland
	\and Institute for Astronomy, Royal Observatory Edinburgh, Blackford Hill, Edinburgh, EH9 3HJ, UK
	\and Armagh Observatory, College Hill, Armagh BT61 9DG, Northern Ireland, UK
}
       

\date{Received; accepted }

\abstract{}
{We present an analysis of a peculiar supergiant B-type star (VFTS698/Melnick\,2/Parker\,1797) in the 30 Doradus region of the Large Magellanic Cloud 
which exhibits characteristics similar to the broad class of B[e] stars. }
{We analyse optical spectra from the VLT-FLAMES survey, together with archival optical and infrared photometry and X-ray imaging to characterise the system.}
{We find radial velocity variations of around 400\,\kms in the high excitation \ion{Si}{iv}, \ion{N}{iii} and \ion{He}{ii} spectra, and photometric variability of 
$\sim$0.6\,mag with a period of 12.7\,d.  In addition, we detect long-term photometric variations of $\sim$0.25\,mag, which may be due to a longer-term 
variability with a period of $\sim$400\,d.}
{We conclude that VFTS698 is likely an interacting binary comprising an early B-type star secondary orbiting a veiled, more massive companion.  
Spectral evidence suggests a mid-to-late B-type primary, but this may originate from an optically-thick accretion disc directly surrounding the primary.}



\authorrunning{P.~R.~Dunstall et al.}
\titlerunning{A peculiar B[e]-like supergiant in 30~Doradus}

\maketitle
%
\section{Introduction}                                         \label{s_intro}

Massive stars, whilst intrinsically rare due to the slope of the initial mass function \citep{sal55}, have an important role in influencing their surroundings.  As the progenitors of core-collapse supernovae \citep[e.g.][]{sma09}, they inject both nuclear processed material and energy into the interstellar medium.  Additionally their high luminosity permit us to observe them beyond the Milky Way \citep[e.g.][]{mas03,lev05,mas09}, providing a crucial test of stellar evolution at different metallicities.


Variability is a defining characteristic for several of the subclasses of massive stars, e.g., luminous blue variables (LBVs) and Cepheids. Such variability can take a wide variety of forms: pulsations in Cepheids can give rise to stable, periodic light-curves, while LBVs undergo irregular and unpredictable eruptive outbursts \citep[e.g. the most famous, Eta Carina,][]{gen84} which can lead to sudden increases in magnitude. Individual types of massive stars can also display more than one type of variability, for example some LBVs can also display gradual changes in magnitude over timescales of years.

Spectroscopic and photometric variability in massive stars can also be caused by binarity, with as many as 40\% of massive stars found to have binary companions (B-type: Dunstall et al.  to be submitted, O-type: Sana et al. to be submitted). \citet{tar00} presented a review of short-period interacting binaries which give rise to emission-line phenomena during the course of their evolution. Short-period variability ($\sim$5 d) in emission features is associated with ``classical Algol'' objects, whereas variability of the order of tens of days can be seen in W~Serpentis binaries or double periodic variables \citep[DPVs,][]{pla80,men03}.

B[e] supergiants (sgB[e]) are the most homogeneous and distinct group of stars that show the `[e]' phenomenon \citep{lam98}.  Their spectra are characterised by the presence of narrow emission from both permitted and forbidden lines of iron-group elements. SgB[e] stars have high luminosities (log~$L/L_{\odot}$\,$>$\,4) and are therefore easily observable in the Magellanic Clouds. Ten sgB[e]s in the Large Magellanic Cloud (LMC) are well documented\footnote{\citet{lam98} listed 11 B[e] supergiants in the LMC, later revised to 10 by \citet{bon09}.}, allowing concise criteria to be placed upon the characterisation of these stars \citep{lam98}.  They show a possible evolutionary connection to LBVs, e.g., \citet{her10}, defined V39 in the Small Magellanic Cloud (SMC) as either an LBV candidate or a sgB[e].  In addition, sgB[e]s may be the result of binary mergers \citep{pod06}, although their exact nature remains unclear at present.

In this paper we document the spectral and photospheric characteristics of an intriguing radial velocity (RV) variable from
the VLT-FLAMES Tarantula Survey \citep[][VFTS; hereafter Paper~I]{eva11}.  The target, VFTS698, has previously been known as
Melnick\,2 \citep{mel85} and Parker\,1797 \citep{par93}.  Section \ref{s_obs} presents the optical spectroscopy of VFTS698, as
well as archival optical, infrared (IR) and X-ray imaging. The observational characteristics of VFTS698 are discussed in Sect.
\ref{s_analysis}, in addition to a preliminary analysis.  VFTS698 has previously been characterised as a B3 Ia star by \citet{wal97} and
``early Be'' by \citet{bos99}, but in Sect. \ref{s_classification} we present our own classification based on the properties determined
here.  Section \ref{s_stellar} documents the associated atmospheric parameters, including an estimate of the nitrogen abundance.  We 
compare VFTS698 with a number of other sgB[e] stars and interacting binary objects in Sect. \ref{s_comparison}, as
well as presenting a plausible evolutionary scenario.  Our conclusions are drawn up in Sect. \ref{s_conclusions}.

\section{Observations}                                        \label{s_obs}
\subsection{Optical spectroscopy}		\label{s_spectra}

Spectroscopy of VFTS698 was obtained as part of the Tarantula Survey, using the Fibre Large Array Multi-Element Spectrograph
(FLAMES) on the Very Large Telescope (VLT) with the Giraffe spectrograph.  The star was observed as part of `Field~B' of the survey, 
for which the observational epochs are listed in the Appendix of Paper~I.  Three of Giraffe wavelength settings were used, as summarised in Table~\ref{t_obs}. 

The data reduction was discussed at length in Paper~I. In brief, the data were reduced using the European Southern Observatory (ESO) Common Pipeline Library FLAMES package, and consisted of flat-fielding, bias subtraction and wavelength calibration.  Separate fibres were used to observe the sky in each exposure, with the average sky spectrum then subtracted from each of the science targets. However, as discussed in Paper~I, this often leads to imperfect nebular subtraction.  In particular, VFTS698 is embedded in a \ion{H}{ii} region with strong nebular emission lines which vary over small spatial scales, so accurate nebular subtraction was not possible.

Cross-talk between fibres on the array can be a source of contamination between objects (see Paper~I). This was investigated for
VFTS698 using the LR03 frames.  An increase of less than 1\% in the typical {\it inter-fibre} flux was found between VFTS698 and one of the adjacent stars on the array (VFTS591).  Such a small contribution to the flux of VFTS698 should not significantly affect our analysis.

\begin{table}
\caption{Observational spectroscopy for VFTS698.  The number of epochs associated with each setting is given in parentheses.}
\label{t_obs}
\begin{center}
\begin{tabular}{lccc}
\hline\hline
Giraffe set-up & $\lambda$-range  & $R$ & No. Exposures \\
 & (\AA) & & \\
\hline
LR02 & 3960 -- 4560 & $\phantom{1}$7,000 	& 15 (5)\\
LR03 & 4505 -- 5050 & $\phantom{1}$8,500	& $\phantom{1}$6 (1) \\
HR15N & 6470 -- 6790 & 16,000 & $\phantom{1}$4 (1) \\
\hline
\end{tabular}
\end{center}
\end{table}

The extracted spectra were normalised over the entire spectral region for each setting, or, for features of particular interest,
in smaller wavelength regions.  On occasions the richness of the spectra make it difficult to identify line-free regions -- the
implications of this for the silicon spectra are discussed in Sect.~\ref{s_stellar}.  Individual exposures were normalised using
low-order polynomials, with a sigma-clipping algorithm used to exclude cosmic-rays within defined continuum windows.  Spectra within an epoch
and, where appropriate, between epochs were combined using a weighted average together with sigma-clipping. Adopting a median
spectrum leads to very similar results indicating that the choice of algorithm for combining individual exposures was unlikely to be a
significant source of error.


\subsection{Photometry}		\label{s_photometry}

\begin{figure}[hbtp]
\begin{center}
\includegraphics[bb=0 0 581 777,angle = 90,scale=0.33]{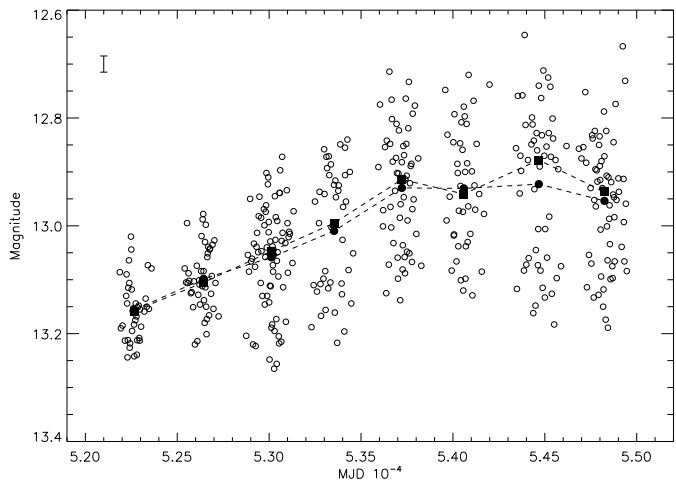}
\caption{$I$-band light-curve of OGLE-III data from 8 years of observations.  VFTS698 shows a short-term variation of $\sim$0.2\,mag, increasing up to $\sim$0.5\,mag with time.  Solid circles and squares show the mean and median values, respectively, for each block.  The typical uncertainty associated with the OGLE-III data is $\pm$0.015\,mag, as illustrated in the upper left of the figure.}
\label{f_light_curve}
\end{center}
\end{figure}

Photometry of VFTS698 is available from the Optical Gravitational
Lensing Experiment III (OGLE-III) survey \citep[][and references
therein]{uda08}, comprising multi-epoch observations over eight
seasons from October 2001 to April 2009. $I$-band photometry for
VFTS698 was obtained from the OGLE database \citep[see][for further
details]{uda03} and is shown in Fig. \ref{f_light_curve}, in which the mean and median measurements for
each observing season are over-plotted.  Note that most of the FLAMES spectroscopy
is contemporaneous with the last block of OGLE-III data.


We have supplemented the OGLE-III photometry with $V$-band observations from the 2-m Faulkes Telescope South (FTS), obtained as part of a monitoring programme of 30~Dor during 2009 and 2010.  The observations employed the FTS Merope camera, which has a field-of-view of 4\farcm7\,$\times$\,4\farcm7 and a (rebinned) pixel size of
0\farcs278/pixel.  Images obtained in seeing of greater than 2$''$ were rejected, leaving 39 observations of the field including VFTS698.  Differential aperture photometry was obtained for VFTS698 via comparison with five (non-variable) stars of comparable brightness in the same field.



To investigate its near-IR behaviour, the {\it Spitzer} Archive\footnote{http://irsa.ipac.caltech.edu/} was searched for
photometry of VFTS698 and nearby sources with comparable $V$-band magnitudes. 
Observations with the 3.6, 4.5, 5.8 and 8.0$\mu$m filters are available from the {\it Spitzer} Legacy Science Program
\citep[SAGE]{mei06}, with photometry from the survey for known massive stars presented by \citet{bon09}.

Four B-type stars with similar $V$-band magnitudes (including VFTS450, an O9.7 star) are located within 2$'$ of VFTS698.  While VFTS698 was
detected in all filters, this was not the case for the comparison objects.  For common detections, VFTS698 was found to be on average
2\,mag brighter at 3.6$\mu$m, 1.5\,mag at 4.5$\mu$m, 2\,mag at 5.8$\mu$m, and the only object detected at 8.0$\mu$m.  Representative
visual and IR magnitudes are summarised in Table \ref{t_photo}, although we emphasise that VFTS698 is photometrically variable.

\subsection{X-ray data}		\label{s_x_ray}

We searched the Chandra Data Archive\footnote{http://cda.harvard.edu/chaser/} for images of the
region including VFTS698. Archival images with the ACIS-I instrument of this part of 30~Dor are available from 21~Sept 1999, and 21, 22 and 30~Sept 2006. The primary data products from the Chandra X-ray Centre's automated processing were downloaded and examined, but no obvious source was found coincident with VFTS698.  This supports the analysis of the 1999 data by \citet{tow06} who did not report a detection at the position of VFTS698 (given a lower detection limit of 5 counts in 21\,870\,s, which equates to a limiting full-band luminosity of $\sim$1\,x\,10$^{33}$\,ergs\,s$^{-1}$, with the brightest source detected being $<$10$^{36}$\,ergs\,s$^{-1}$).

\begin{figure}[hbtp]
\begin{center}
\includegraphics[angle = 0,scale=.5]{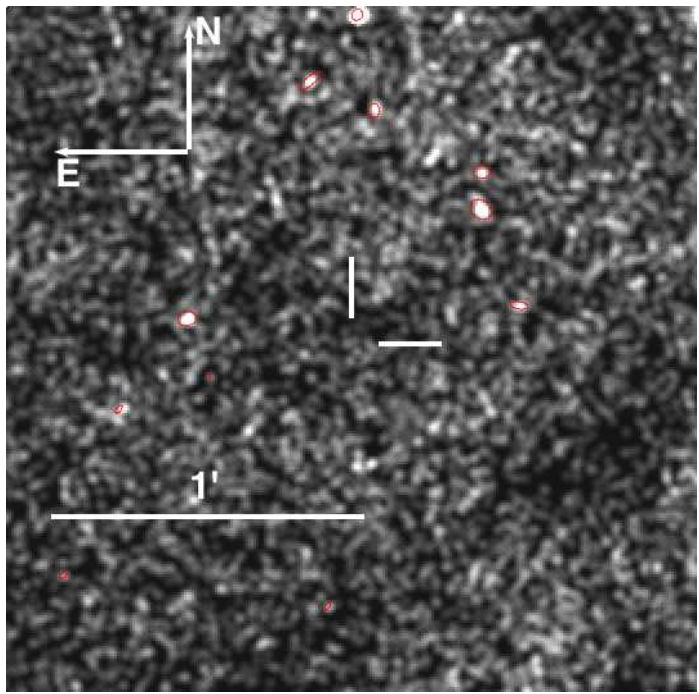}
\caption{Merged Chandra image (smoothed for legibility) of the region surrounding VFTS698 (marked by the white crosshairs).  
}
\label{f_xray}
\end{center}
\end{figure}

We reprojected all of the available individual exposures to a common pointing and stacked them to create a deep image with an
effective exposure time of 120\,ks. The deep, merged image (Fig.~\ref{f_xray}) was filtered to only include photons with energies
in the range of 0.3 to 10 keV and a wavelet algorithm was used to detect point sources. Even when an extremely low significance threshold
was used (corresponding to a false positive rate among the detections of 20\%), no source was found to be coincident with, or close to,
VFTS698.

\section{Observational properties and preliminary analysis}                              \label{s_analysis}

\subsection{Optical photometry}                              \label{s_photo_analysis}

From Fig. \ref{f_light_curve} there is clear evidence for photometric variability of $\sim$0.6\,mag on timescales ranging from days to years. Two separate trends can be identified: short-term variability with a period of 12.7\,$\pm$\,0.1\,d, and a longer-term variation with timescales of at least several hundred days.

\begin{figure*}
\begin{center}$
\begin{array}{cc}
\includegraphics[scale=0.43]{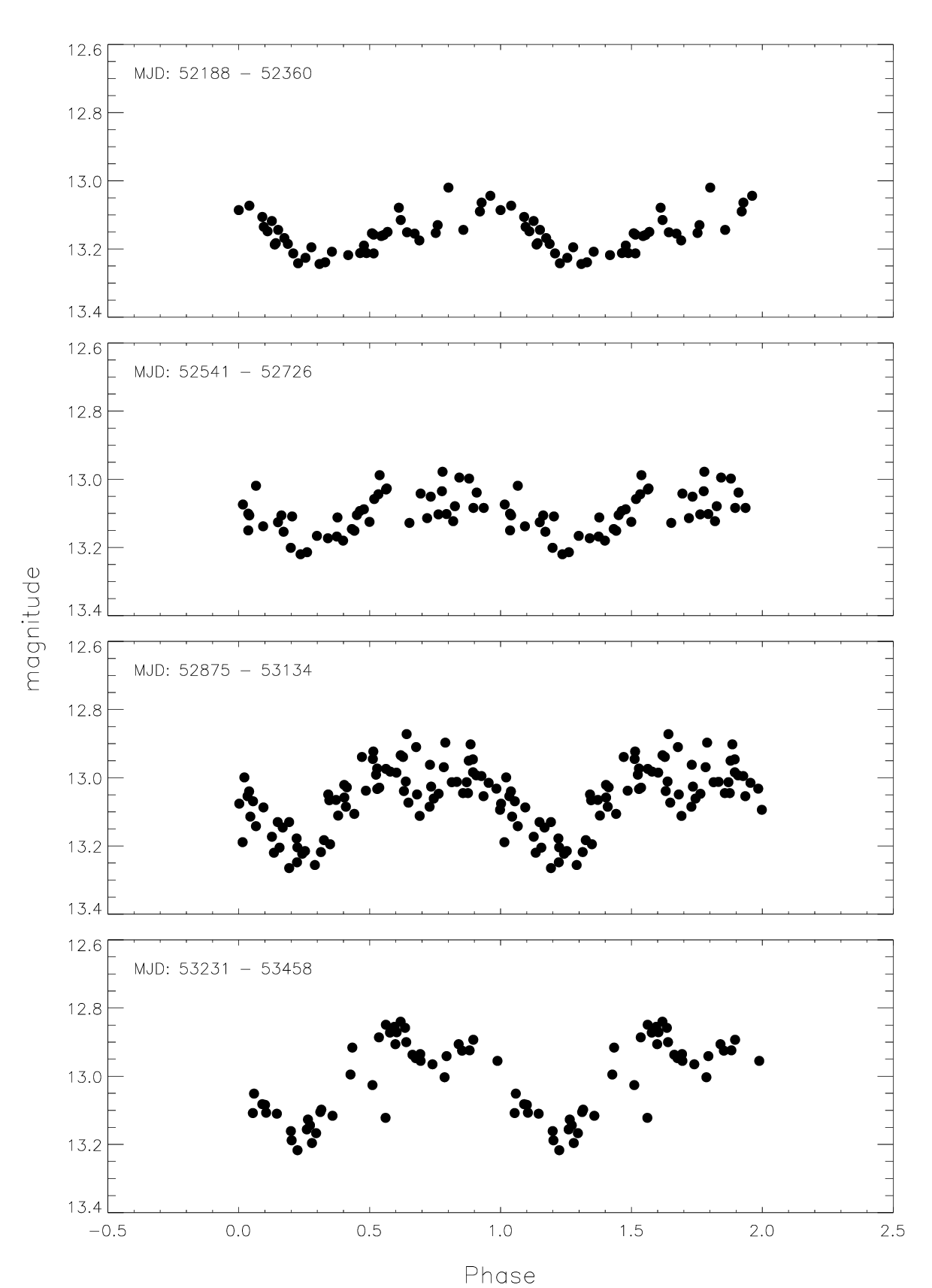} & \includegraphics[scale=0.43]{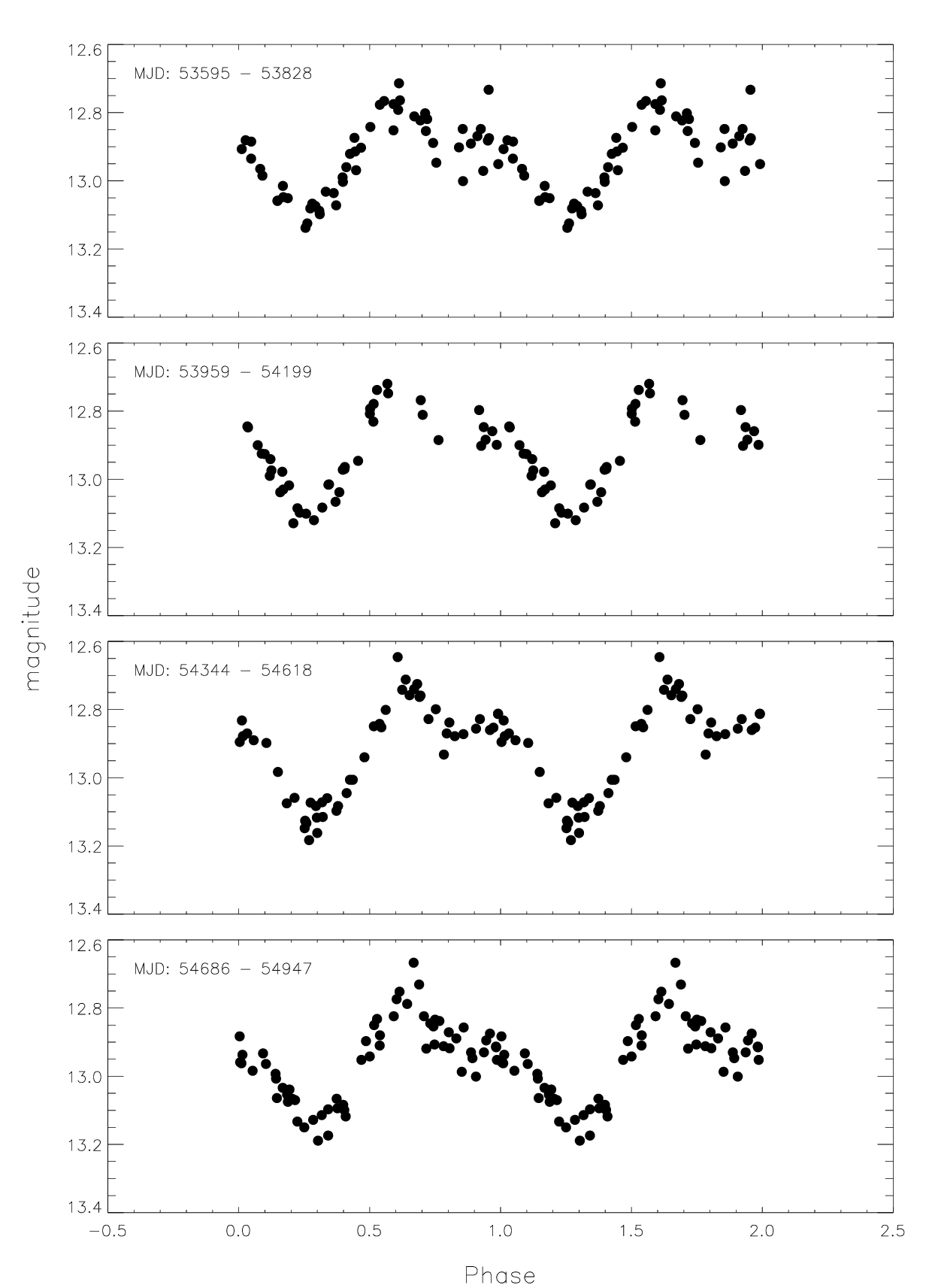} \\
\end{array}$
\caption{OGLE-III $I$-band photometry of VFTS698 folded with a 12.7\,d period.  {\it Left-hand panel:} First four blocks of observations; 
{\it Right-hand panel:} blocks five to eight.  For clarity two periods have been plotted for each observation block.}
\label{f_phased_light}
\end{center}
\end{figure*}

The short-term variation is illustrated in  Fig.~\ref{f_phased_light}. Within the observational uncertainty the period appears to be constant between epochs, but the amplitude of the periodicity increases in magnitude over the 8 years, from $\sim$0.10 to $\sim$0.25\,mag.  Each observation block was analysed individually by means of a Lomb-Scargle periodogram, confirming the consistency of the 12.7\,d period.    

\begin{figure*}
\begin{center}$
\begin{array}{cc}
\includegraphics[bb=0 0 561 767, angle=90, scale=0.34]{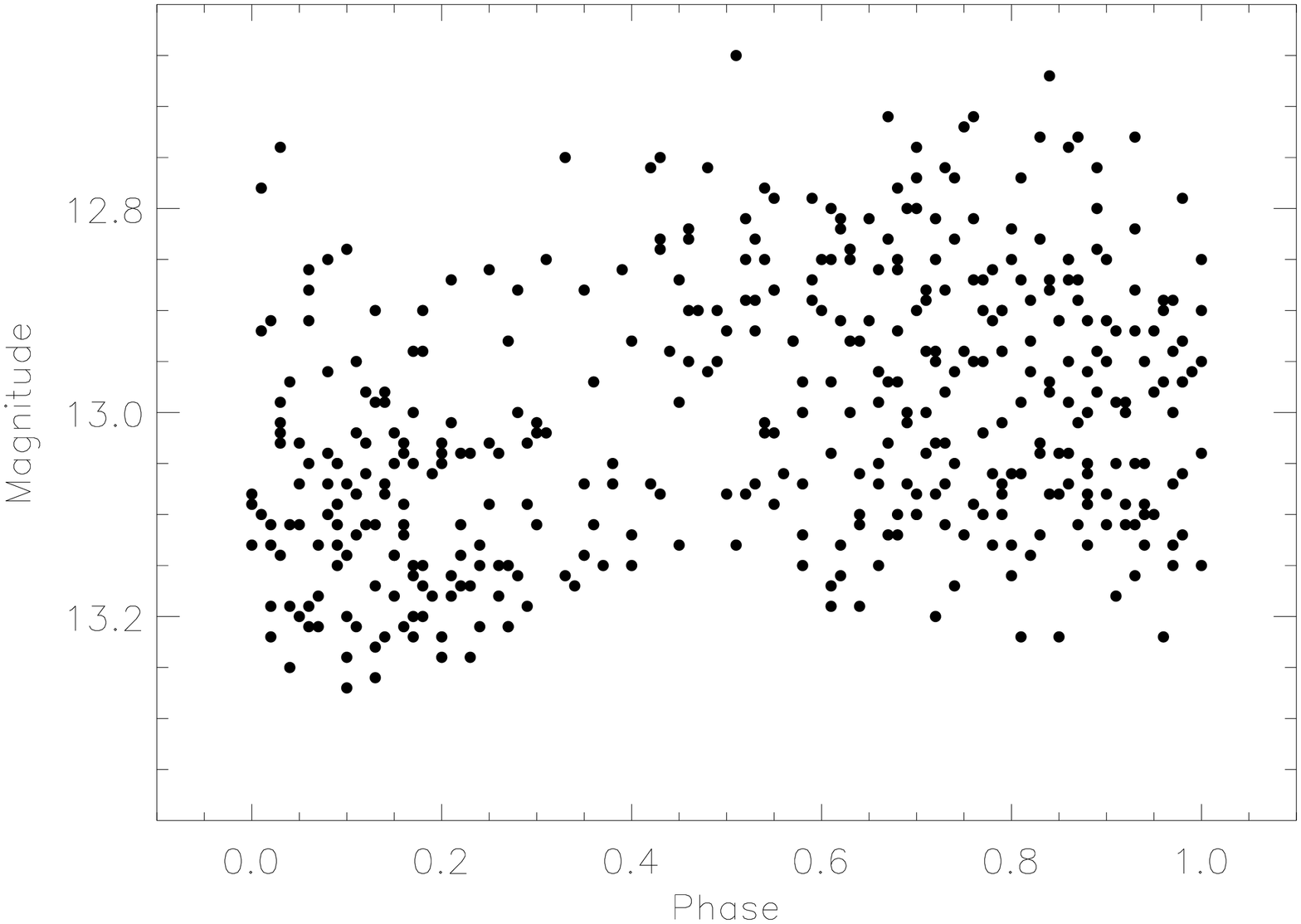} & \includegraphics[bb=0 0 561 767, angle=90, scale=0.34]{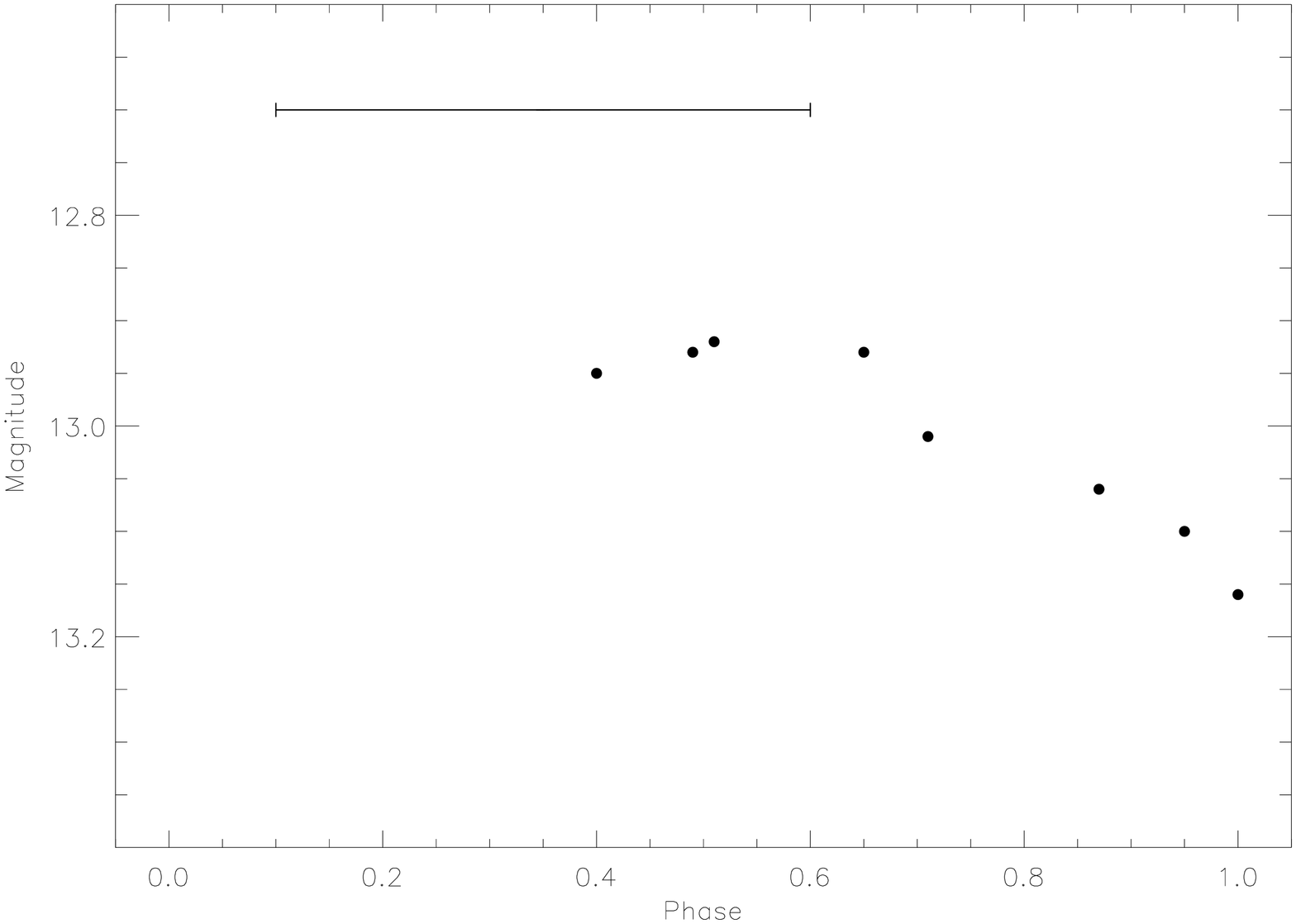}\\
\end{array}$
\caption{ {\it Left-hand panel:} $I$-band OGLE-III photometry of
  VFTS698 folded to a 400\,d period. {\it Right-hand panel:} The
  median magnitude and date for each block of OGLE-III $I$-band data
  as a function of phase. The bar indicates the typical phase coverage
  by each block.}
\label{f_400}
\end{center}
\end{figure*}

The long-term trend appears to consist of a steady increase in the mean $I$-band magnitude of $\sim$0.3\,mag over a period of some years, which may have reached a maximum at MJD $\sim$54500 (see Fig.~\ref{f_light_curve}).  To investigate any possible periodicity in this trend, a Lomb-Scargle periodogram was produced from the $I$-band photometry, but now restricted to a period range of 100 to 1000\,d.  The result of this indicated a period of 400 $\pm$ 21\,d, but upon folding the period to the complete $I$-band observations, no periodic phase is apparent (left-hand panel of Fig.~\ref{f_400}).  The difficulty in identifying any periodicity in this long-term variation is due to the large brightness changes associated with the 12.7\,d period.  When the mean magnitudes for each block are plotted against the median phase a more convincing variation is found, as shown by the right-hand panel of Fig.~\ref{f_400}.


\begin{figure}[hbtp]
\begin{center}
\includegraphics[angle=90,scale=.3]{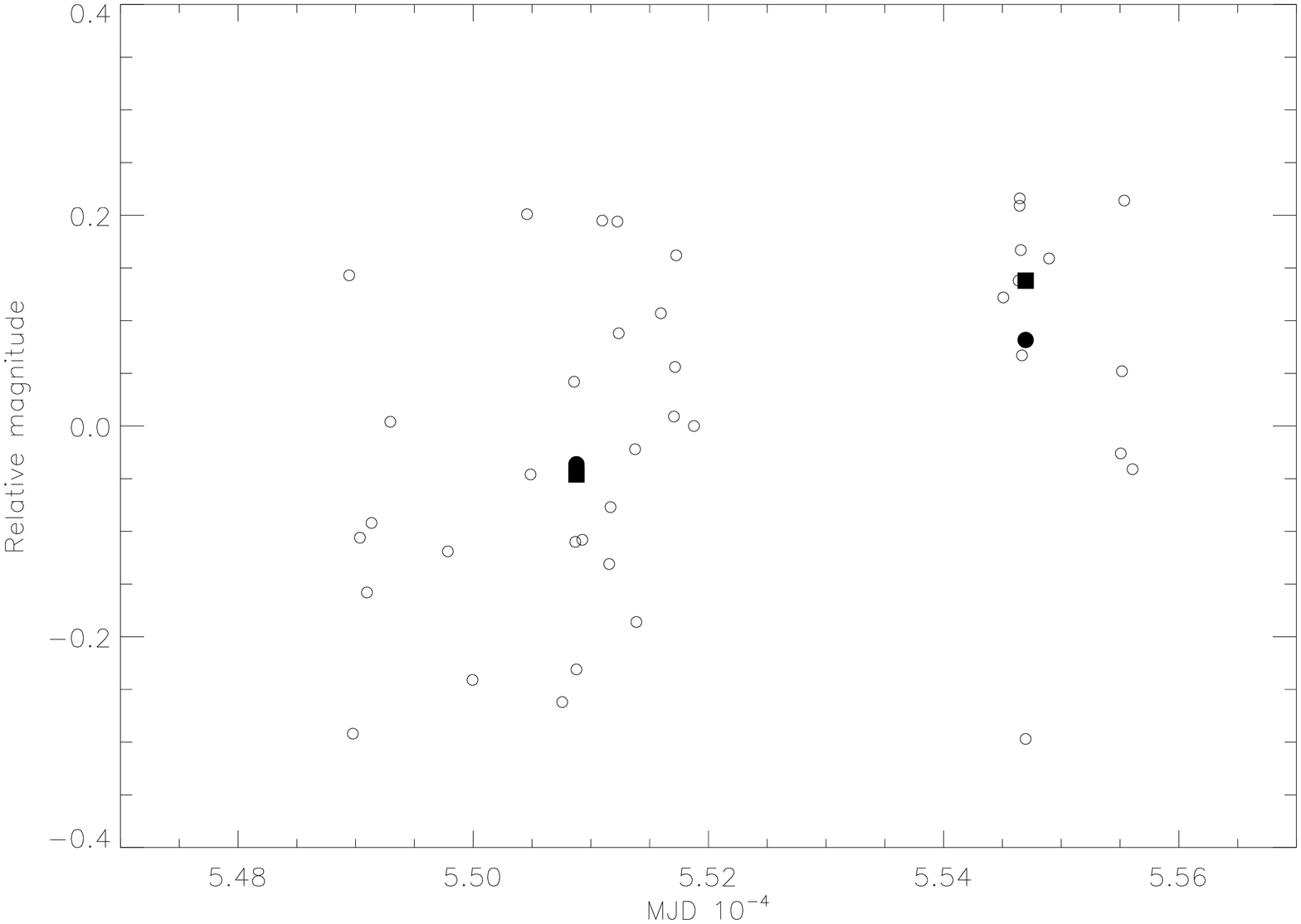}
\includegraphics[angle=90,scale=.3]{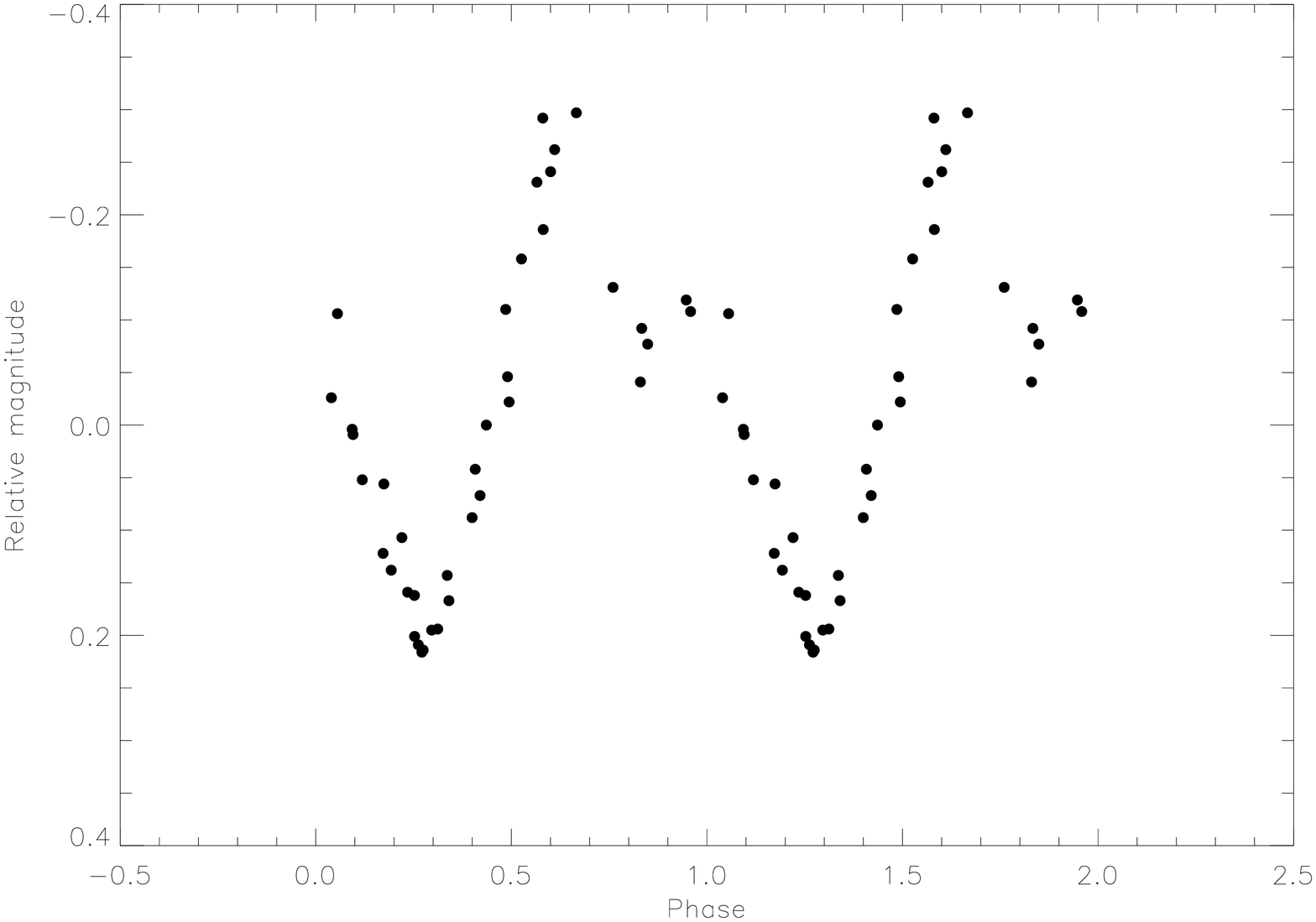}
  \caption{{\it Upper panel:} Differential $V$-band photometry of
    VFTS698 from 11 months of FTS observations (open circles).  Mean
    and median magnitudes for the two seasons are indicated by solid
    circles and squares, respectively. {\it Lower panel:} FTS
    photometry folded to a 12.7\,d period.}
  \label{f_phased_faulkes}
\end{center}
\end{figure}

As discussed in Sect.~\ref{s_photometry}, VFTS698 was also observed over an 11 month period in the $V$-band with the FTS.  These data were found 
to be in good agreement with the OGLE-III data; a Lomb-Scargle periodogram yielded a period of 12.7\,$\pm$\,0.1\,d (hereafter referred to as the 
12.7\,d period) with a photometric amplitude of $\sim$0.25\,mag. The phased light-curve from the FTS data is shown in Fig.~\ref{f_phased_faulkes}.

\subsection{Infrared photometry}	\label{s_infra_analysis}

\begin{figure}[hbtp]
\begin{center}
\includegraphics[bb=0 0 581 767,angle=90,scale=.34]{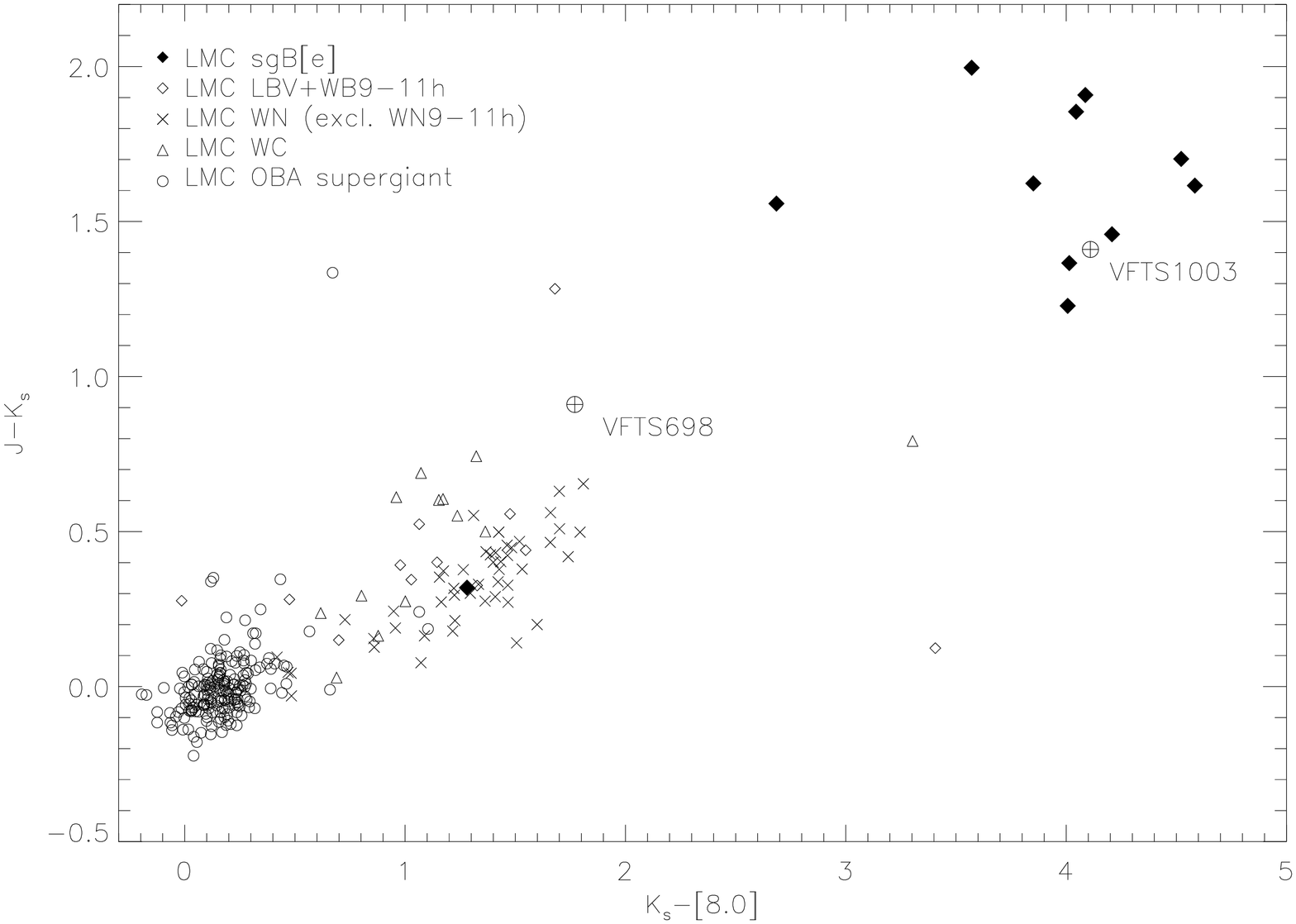}
  \caption{$J-K_s$ colour vs. $K_s-$[8.0] colour for VFTS698, other
    emission-line objects and OBA-type supergiants in the LMC.  The
    recently discovered B[e] star VFTS1003 is also shown for
    comparison. Note that strong emission lines are also expected to
    contribute to the near-IR continuum emission of WC stars.} \label{f_IR_near}
\end{center}
\end{figure}

\begin{figure}[hbtp]
\begin{center}
\includegraphics[bb=0 0 581 767,angle=90,scale=.34]{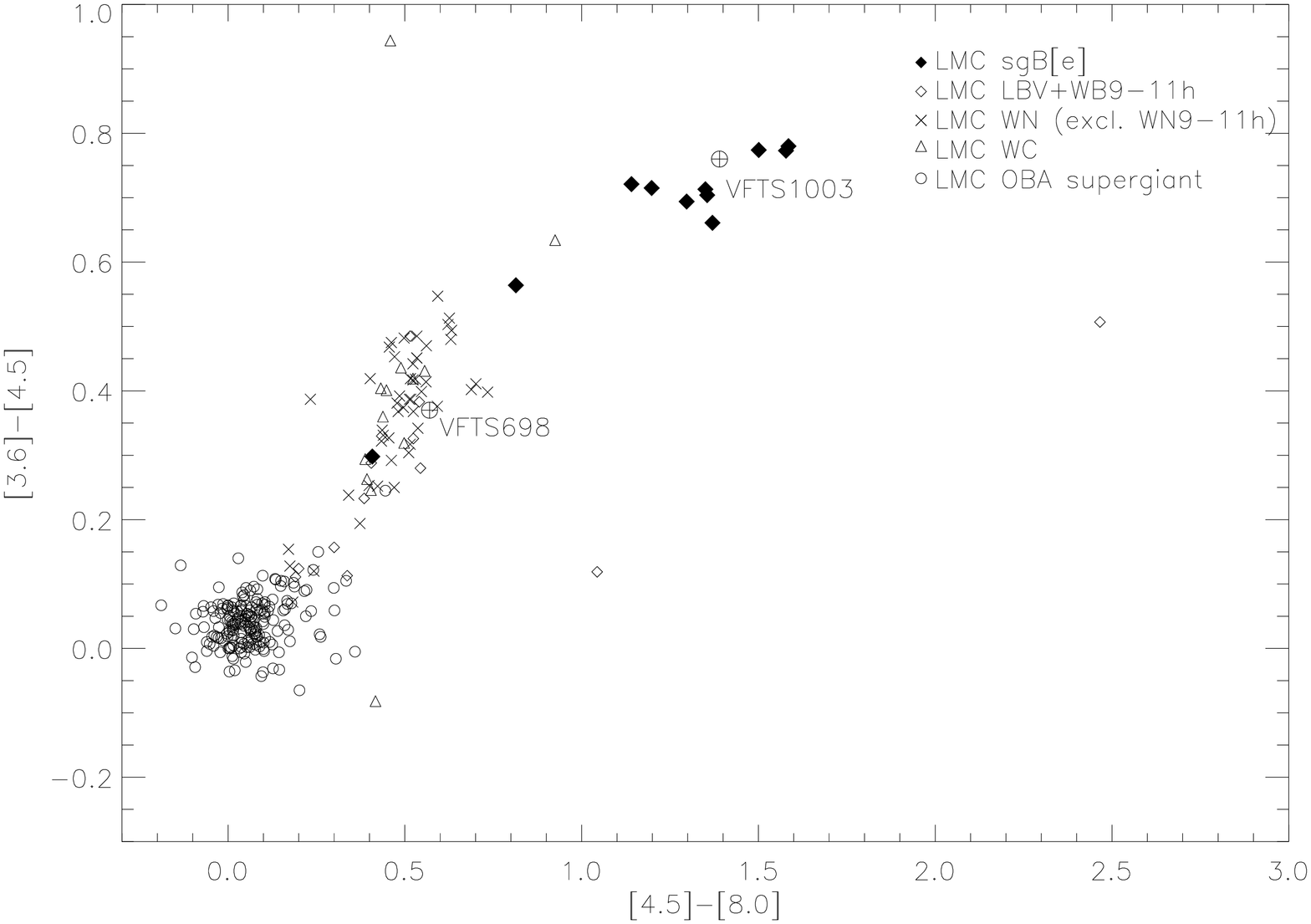}
  \caption{Mid-IR colours for VFTS698, other emission-line objects
    and OBA-type supergiants in the LMC.  VFTS698 appears to have an
    IR-excess greater than normal OBA supergiants, but not to the same
    extent as other sgB[e] stars.} \label{f_IR}
\end{center}
\end{figure}

The near- and mid-IR colours of VFTS698 are compared to other massive stars in the LMC in Figs. \ref{f_IR_near} and \ref{f_IR}; near-IR photometry is from the InfraRed Survey Facility \citep[IRSF,][]{kat07} and longer wavelengths are from the {\it Spitzer} SAGE survey.  To investigate its behaviour, we also include results for VFTS1003, the B[e] star discussed in Paper~I.  As shown in the 
figures, both VFTS698 and VFTS1003 display significant IR excesses over `normal' OBA-type supergiants.  

VFTS698 is clearly displaced from the region of the colour/colour plots occupied by sgB[e] stars.  At mid-IR wavelengths it is
co-located with other massive stars such as LBVs and Wolf-Rayet (WR) stars, which are known to support strong dense winds resulting in copious thermal free-free emission at near- to mid-IR wavelengths\footnote{We note that the source of such excesses in the near-IR for LBVs may also be small ejecta nebulae surrounding the star \citep{bon09}.}.  The near-IR colours of VFTS698 are bluer than {\em bona fide} sgB[e] stars, but they indicate a greater continuum excess than found for either the LBVs or WRs.  The origin of the IR continuum emission in VFTS698 is therefore uncertain; in Sect.~\ref{s_comparison} we argue that it appears likely to arise (at least partially) in an ionised stellar wind or gaseous circumstellar/binary disc, although an additional component from hot dust would also appear possible.

In contrast, we note that VFTS1003 is co-located with the sgB[e] stars in Figs.~\ref{f_IR_near} and \ref{f_IR}, suggesting the presence of a warm ($\sim$600\,K, see Sect.~\ref{s_comparison}) circumstellar dust.  
Note that there is some ambiguity regarding the classification of VFTS1003 in Paper~I, with it potentially a sub-luminous sgB[e] or an over luminous, pre-main sequence `Herbig' B[e] star.  The near- and mid-IR photometry would appear to support the sgB[e] classification, but the close proximity of VFTS1003 to the central region of 30-Dor casts doubt on the reliability of the Spitzer photometry.

\begin{table*}
\caption{Photometry of VFTS698 and four nearby, early B-type (or late O) VFTS sources with comparable $V$-band magnitudes. Photometry
of VFTS1003, provisionally classified as B[e] in Paper~I, is also included for comparison.}
\label{t_photo}
\begin{center}
\begin{tabular}{llllccccccccc}
\hline\hline
Star		& Alias & Lit. & VFTS & $V$ & $B - V$  & $J$ & $H$ & $K_s$ & [3.6] & [4.5] & [5.8] & [8.0] \\
 & & Sp. type & Sp. type & & & & & \\
\hline
VFTS450\tablefootmark{\dagger} & Mk50 & ON9: I\tablefootmark{1} & O9.7 & 13.60 & 0.20 & 13.08 & 12.91 & 12.89 & 11.00 & 10.65 & 10.52 & ... \\
VFTS652\tablefootmark{\dagger} & Mk5 & B2 Ib:\tablefootmark{1} & ... & 13.88 & 0.20 & 13.40 & 13.28 & 13.22 & 12.97 & ... & 12.73 & ... \\
VFTS698\tablefootmark{*} & Mk2 & B3 Ia\tablefootmark{1} & ... & 13.68 & 0.44 & 12.34 & 11.89 & 11.43 & 10.60 & 10.23 & 10.02 & 9.66 \\
VFTS732\tablefootmark{*} & Mk1 & ... & ... & 13.03 & 0.19 & 12.35 & 12.19 & 12.08 & 11.85 & 11.73 & 11.30 & ... \\
VFTS733\tablefootmark{*} & P1988 & B0.5 V\tablefootmark{2} & ... & 14.28 & 0.12 & 13.90 & 13.82 & 13.74 & 13.59 & ... & ... & ... 
\\
VFTS1003\tablefootmark{\dagger} & S99-283 & ... & B[e]? & 16.10 & 0.23 &13.86 & 13.32 & 12.45 & 10.49 & 9.73 & 9.02 & 8.34  \\
\hline
\end{tabular}
\tablefoot{Spectral types from: \tablefoottext{1}{\citet{wal97}};\tablefoottext{2}{\cite{bos99}}.
Optical photometry from: \tablefoottext{*}{\citet{par93}}; \tablefoottext{\dagger}{\citet{sel99}}.
Near-IR ($JHK_s$) photometry from IRSF Magellanic Clouds catalogue \citep{kat07}.  Mid-IR photometry taken from Spitzer SAGE \citep{mei06}.}

\end{center}
\end{table*}




\subsection{Spectral classification and luminosity}	\label{s_classification}

\begin{figure}[hbtp]
\begin{center}
\includegraphics[bb = 0 0 581 767,angle = 90,scale=0.34]{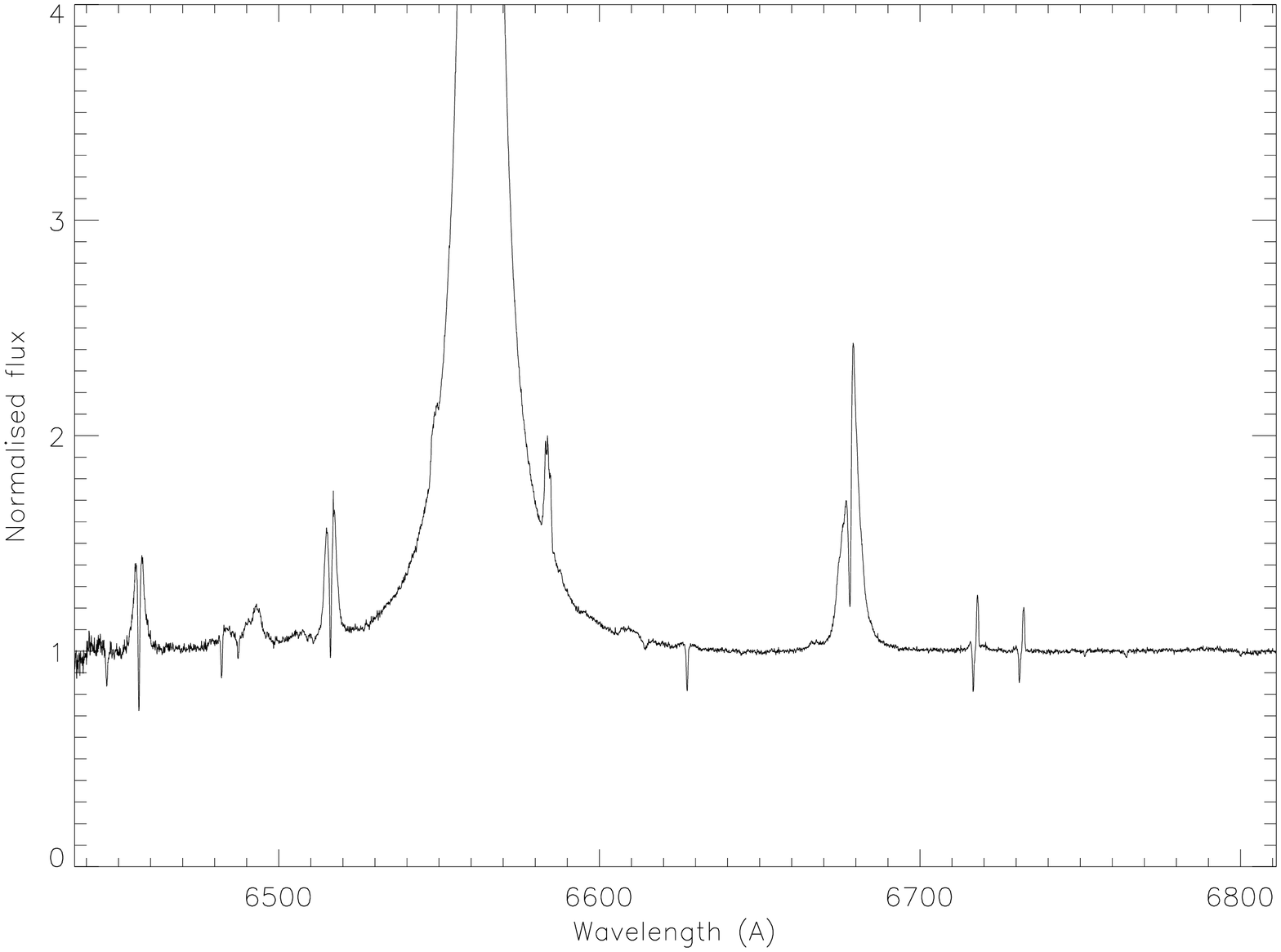}
  \caption{Optical spectroscopy of the HR15N observations of VFTS698.  See Fig. \ref{f_698_R4_R50} for details on the LR02 and LR03 settings.
}
\label{f_spectra}
\end{center}
\end{figure}

VFTS698 shows a rich absorption- and emission-line spectrum, as illustrated in Figs. \ref{f_spectra} and \ref{f_698_R4_R50}.  \citet{lam98} presented an
overview of the sub-classification of B[e] stars, from which we classify VFTS698 as a B[e] supergiant.  Our two primary criteria were
the presence of forbidden emission lines and a large luminosity.
In addition, five secondary criteria were considered: indications of mass-loss from double-peaked Balmer emission lines; presence of
narrow, low-excitation emission lines and broad, higher-excitation absorption lines; a strong nitrogen spectrum; a large extinction;
photometric variability.  For the last point, VFTS698 shows larger variations than most sgB[e] objects (typically 0.1 to
0.2\,mag), but it is not unique -- e.g. R4 in the SMC displays comparable variability \citep{zic96}. The
strong photometric variability and mid-IR colours appear to weaken the sgB[e] classification, but there remains sufficient evidence to
justify the classification as an early-type supergiant with `[e]' phenomena. 


The photometric variability of VFTS698 makes its luminosity difficult to estimate. However we can use a number of constraints to
obtain a best estimate.  Using the photometry from Table \ref{t_photo}, adopting $(B-V)_0$\,$=$\,$-$0.24\,mag and a distance
modulus of 18.5 \citep{alv04} to the LMC, we obtain $E(B-V)$\,$=$\,0.6\,mag and $M_{V}$\,=\,$-$6.7\,mag.  Adopting an
appropriate bolometric correction for an early B-type supergiant \citep[$-$2.4\,mag, see][]{cro06} results in a bolometric magnitude of $-$9.1\,mag, equivalent to log~$L/L_{\odot}$\,=\,5.6. 

\subsection{Spectral characteristics}	\label{s_lines}

We now discuss the FLAMES spectra, which appear to have three distinct components:
\begin{itemize}
\item{{\it Characteristic sgB[e] spectrum:} Low-excitation permitted and forbidden lines from iron-group elements;}
\item{{\it Cool component:} Absorption lines from species consistent with a mid-B spectral classification, e.g., \ion{He}{i} and \ion{Si}{ii};}
\item{{\it Hot component:} High-ionisation stages (\ion{N}{iii}, \ion{Si}{iv} and \ion{He}{ii}), consistent with an early B- or late O-type spectrum.}
\end{itemize}

\begin{figure}[hbtp]
\begin{center}
\includegraphics[bb=0 0 581 767,scale=0.34,angle=90]{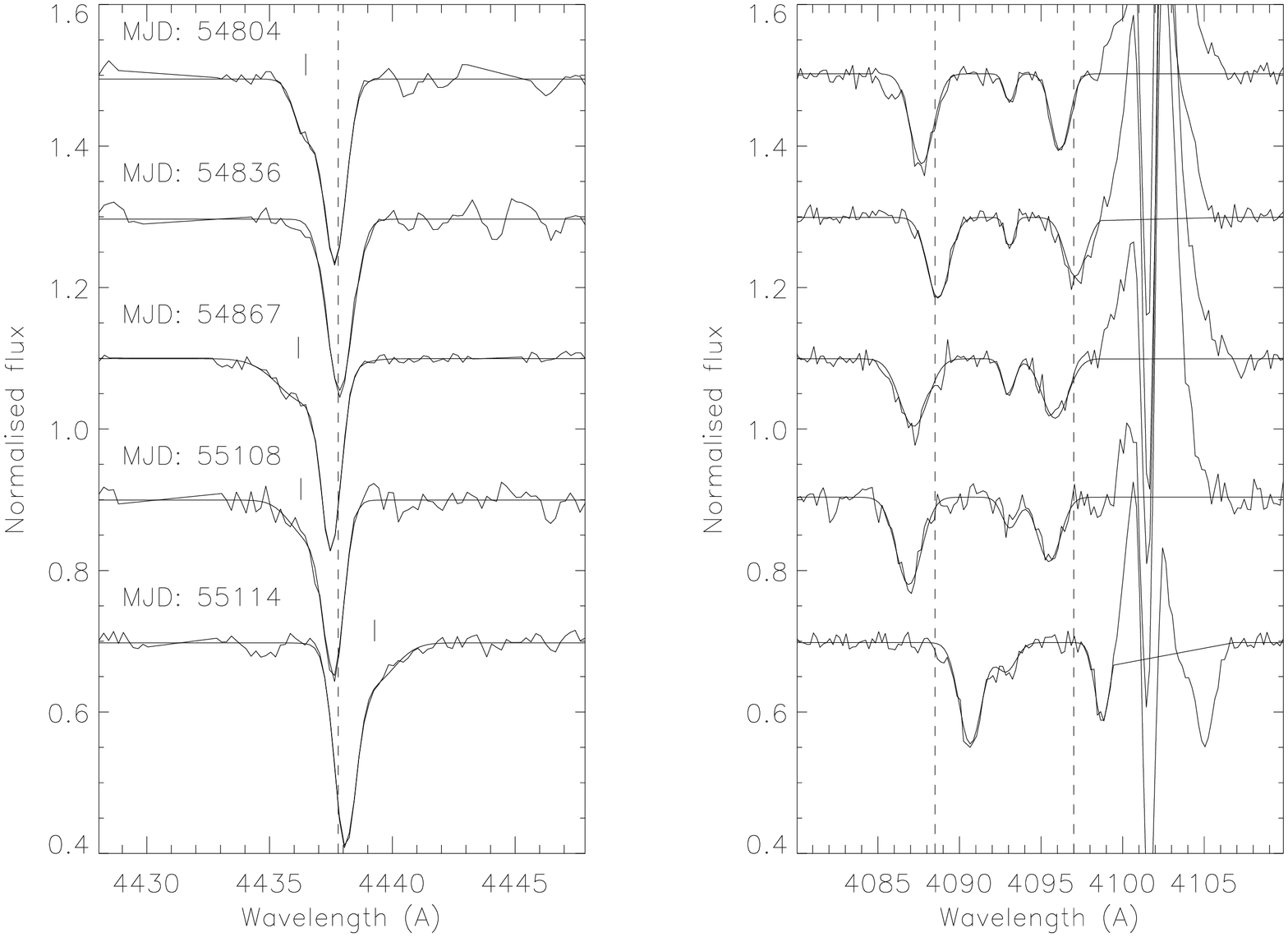}
  \caption{Example RV variations displayed by helium and metal
    lines.  Fitted Gaussian profiles are over-plotted and rest
    wavelengths are indicated by the vertical dashed lines.  {\it
      Left-hand panel:} \ion{He}{i} 4437\AA, in which the `hot', second
    component is noted by the marks above the continuum. {\it
      Right-hand panel:} \ion{Si}{iv} 4089\AA\/ and \ion{N}{iii}
    4097\AA.}
\label{f_si_he}
\end{center}
\end{figure}

The apparently cool features display small RV variations, with far larger variations in the lines attributable to a hot component.
The RV variations of example metal and helium lines (\ion{Si}{iv}, \ion{N}{iii} and \ion{He}{i}) are shown in Fig. \ref{f_si_he}.  We
now consider these different components in turn, although there is inevitably some overlap in the discussion.

\subsubsection{Characteristic sgB[e] spectrum} \label{sp_csm}

\begin{figure}[hbtp]
\begin{center}
\includegraphics[bb=0 0 581 760, angle = 90,scale=0.34]{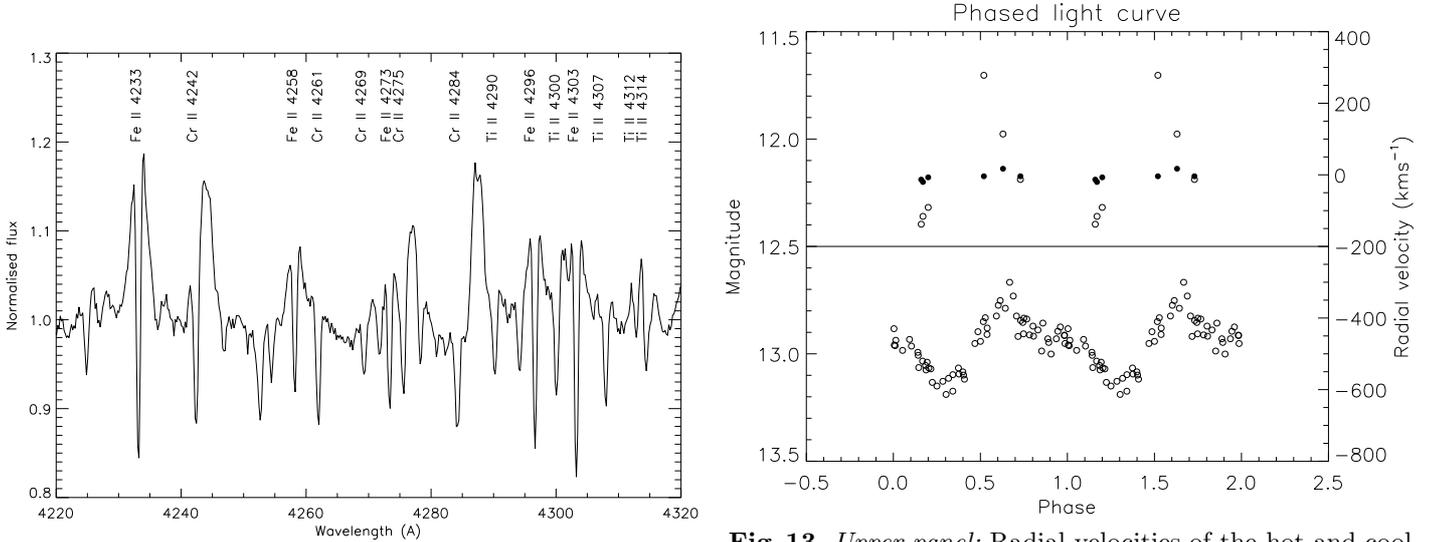}
\caption{\ion{Fe}{ii} , \ion{Cr}{ii} and \ion{Ti}{ii} emission, and shell-like features from the (first epoch) LR02 spectrum of VFTS698.}
\label{f_iron}
\end{center}
\end{figure}

To inform the description of VFTS698, Fig. \ref{f_698_R4_R50} compares the LR02 data with
spectra of two sgB[e] stars in the SMC, obtained with the
Cassegrain Echelle Spectrograph (CASPEC) on the ESO 3.6-m telescope
(degraded and rebinned to match the resolution of FLAMES).
\citet{zic96} noted the prominent emission of the Balmer and
\ion{Fe}{ii} lines (both permitted and forbidden) in R4. Similar
\ion{Fe}{ii} features are seen in VFTS698, but they are accompanied by narrow
absorption lines (see Fig. \ref{f_iron}); such structure is also seen
in the \ion{Ti}{ii} and \ion{Cr}{ii} lines in VFTS698.  This
two-component structure is not seen in R4, but is present in R50
\citep[][upper spectrum in Fig. \ref{f_698_R4_R50}]{zic86}.

The most prominent permitted lines of \ion{Fe}{ii} are due to the multiplets 27, 28, 37 and 38, with an excitation potential $\sim$2.7\,--\,5.6\,eV \citep[similar to those identified by][]{zic96}, whilst for \ion{Ti}{ii} and \ion{Cr}{ii}, they are due to multiplets 41 and 31, with an excitation potential $\sim$1.2\,--\,4.0\,eV and $\sim$3.9\,--\,6.8\,eV, respectively.  The forbidden transitions mainly originate from [\ion{Fe}{ii}] multiplets 4F, 6F, 7F, 20F and 21F, with excitation potentials of 0.1\,--\,3.0\,eV. A number of other forbidden lines are also detected, including [\ion{Fe}{iii}] at 4658, 4701, and 4733\AA, [\ion{N}{ii}] at 6548 and 6583\AA, and [\ion{S}{ii}] at 6717 and 6730\AA.  However, it is unclear if the forbidden transitions of [\ion{N}{ii}] and [\ion{S}{ii}] originate from material directly associated with VFTS698 as these features are also observed in the FLAMES spectra of other nearby objects, where they originate from the strong nebular emission in 30~Dor (Paper~I).

The narrow-lined, iron-group absorption features have been fitted with
Gaussian profiles to estimate RVs.  This was performed for all epochs
of LR02 spectroscopy, and no evidence for RV variations was found, with a mean
RV of 259\,\kms and a sample standard deviation of $\pm$15\,\kms.
This spectral component appears to lie close to the rest frame of
30~Dor, as measured from over two hundred B-type stars
($<$RV$>$\,$=$\,270$\pm$17\,\kms, Kennedy et al. to be submitted).  The low excitation potentials of the iron-group spectrum, together with the RV
measurements which indicate a static environment, suggest they
originate within low density, circumstellar material.


\subsubsection{Cool component} \label{sp_B}

Absorption lines from \ion{N}{ii}, \ion{Si}{ii} and \ion{Mg}{ii} are
observed in the spectrum of VFTS698 with profiles that are broader
than those from the iron-group species discussed above.  There is also
a strong neutral helium spectrum, unusual for objects exhibiting
B[e] features, which show a two-component structure (see Fig. \ref{f_si_he}).

As discussed below the \ion{He}{i} diffuse triplet lines show varying
P~Cygni profiles but the singlet lines appear reasonably symmetric.
Determinations of the RVs of the stronger component of these singlet lines and of
the singly-ionised silicon and magnesium transitions indicate that
they probably originate from the same source.  The \ion{He}{i}
transitions show consistent RV measurements within an
epoch (with a sample standard deviation of $\pm$15\,\kms) but with a
variation of 33\,$\pm$\,11\,\kms over the five spectroscopic epochs. The
\ion{Si}{ii} and \ion{Mg}{ii} lines were subject to blending that made
accurate estimates difficult, but both species were found to have
RVs within typically $\sim$20\,\kms of those of the \ion{He}{i}
lines.  If we consider these lines to originate from a single plasma,
this could be the B[e] supergiant photosphere with a mid-B spectral
classification. Additionally a Lomb-Scargle periodogram for the
\ion{He}{i} RV estimates returned a periodicity of 12.7
$\pm$ 0.1\,d, implying a correlation with the photometric
variability.

\begin{figure*}[hbtp]
\begin{center}
\includegraphics[angle = 90,scale=0.5]{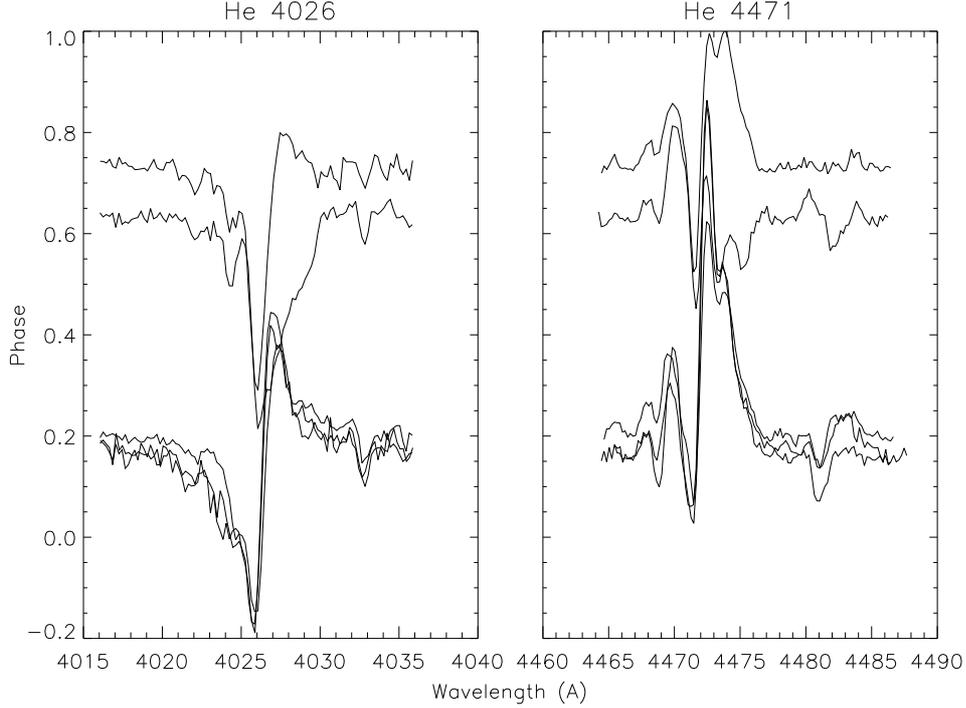}
  \caption{The \ion{He}{i} 4026 and 4471\AA\,normalised line profiles	 (left- and
    right-hand panels, respectively) shifted to their rest
    wavelengths, with their continuum level indicating their phase
    given a single 12.7\,d period.  The three epochs at a phase of
    0.2 display similar structure, whereas later in the 
    period (at phases of 0.6 and 0.8) there are significant changes in morphology.}
\label{f_He}
\end{center}
\end{figure*}

In Fig. \ref{f_He} the strongest \ion{He}{I} diffuse triplet transitions, shifted to their rest wavelength, are shown as a function of their phase, adopting the 12.7\,d photometric period.  It should be noted these features are susceptible to contamination by narrow nebular emission, which may not have been completely removed.  However, variations in the P~Cygni nature of their profiles are clearly visible. Hereafter we refer to these features as the `cool stellar' component.

\subsubsection{Hot component} \label{sp_O}
 
High-ionisation absorption lines were also identified in the spectrum,
viz. \ion{Si}{iii}, \ion{Si}{iv} and \ion{N}{iii} features as well the
strongest \ion{He}{ii} transition in this wavelength region, 4686\AA.
The \ion{Si}{iv} and \ion{N}{iii} features were observed in the multi-epoch
LR02 data, which reveal large RV variations with a range of 
$\sim$260\,\kms. Including the LR03 observation (which includes the
\ion{He}{ii}, and \ion{Si}{iii} lines) this range increases to
$\sim$420\,\kms.  
The weaker component of the \ion{He}{i} spectrum (discussed in the previous
section) has a similar velocity structure to these high-ionisation metal lines. Additionally, the
\ion{N}{ii} spectrum from the LR03 observation also displays a two
component structure, with the RV estimate of one component
compatible with those of the other high-ionisation features.  Although
the weaker \ion{He}{i} component appears to have a similar RV
variation to those of the high-ionisation metal lines, it is normally
blended.  Hence, the velocity range was determined only from the
metal lines.  Assuming that these lines all originate from a single
stellar object, it would have an early-B/late-O spectral classification.  

\begin{figure}
\begin{center}
\includegraphics[bb = 30 300 550 500,angle = 90,scale=0.35]{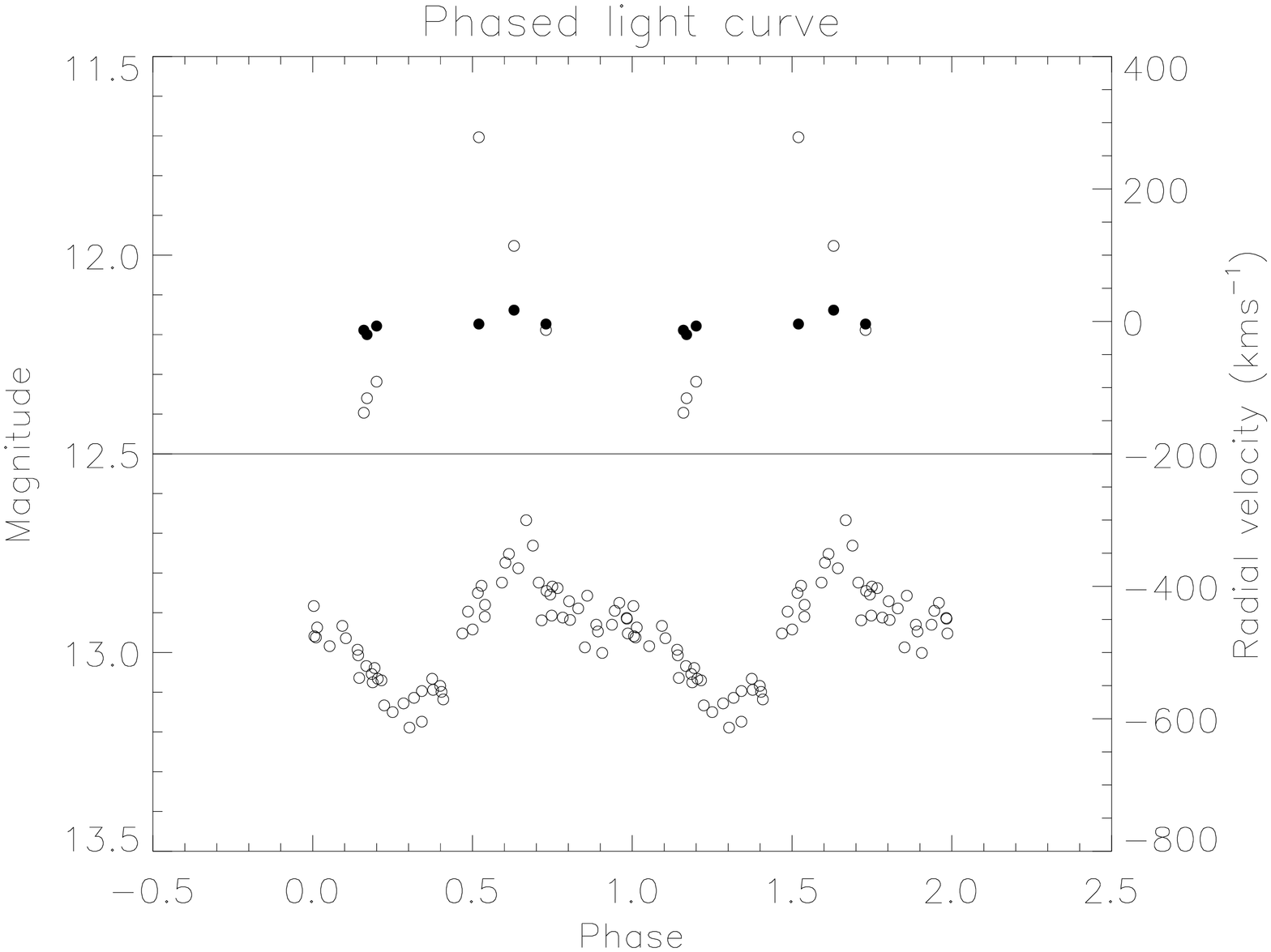}
  \caption{{\it Upper panel:} Radial velocities of the hot and cool components (open and solid circles, respectively) folded with a 12.7\,d period.  {\it Lower panel:} Single block of phased OGLE-III data.  The radial velocity measurements for the hot component appear to be consistent with the periodicity of the photometric data.}
\label{f_rv_mag}
\end{center}
\end{figure}

In Fig. \ref{f_rv_mag} the RV estimates from the high-ionisation
features are folded with the 12.7\,d period found from the OGLE-III data.
Also shown are the OGLE-III data for the block that overlapped in time with most
of the FLAMES spectroscopy.  From the figure the RV data appear to be
consistent with the variations in magnitude and, although there are
insufficient points to obtain an accurate period, a Lomb-Scargle
periodogram returns a period of 10.5 $\pm$ 0.1\,d, which appears
consistent with the photometry.

A detailed list of the absorption and emission features for VFTS698
are presented in Table \ref{t_lines} (available online).  Values for the
central wavelengths (corrected for the RV estimated from the iron-group
spectra), equivalent widths, and full-width half-maxima of the profiles are 
provided.  For spectral lines observed with the LR02 setting, these
measurements are from the first epoch.

\begin{figure*}[hbtp]
\begin{center}
\includegraphics[angle = 90,scale=0.6]{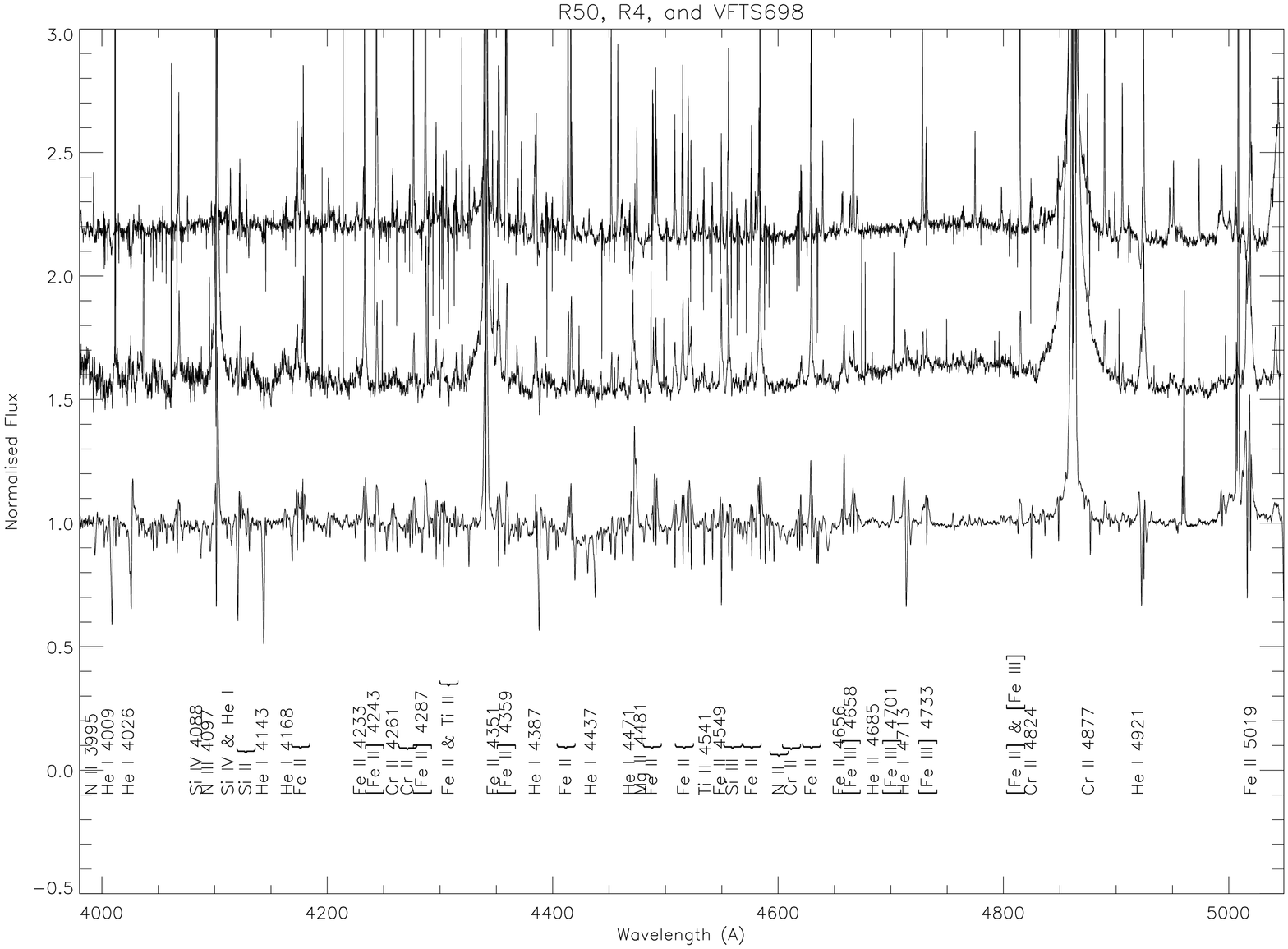}  
  \caption{The combined LR02 and LR03 spectra of VFTS698 (lower spectrum), compared with two sgB[e] stars in the SMC:
  R50 \citep[upper spectrum, from][]{zic86} and R4 \citep[middle spectrum, from][]{zic96,pod06}.  
}
\label{f_698_R4_R50}
\end{center}
\end{figure*}

\begin{table*}
\begin{center}
\caption{Radial velocity measurements of absorption features for selected transitions given for both hot and cool components.  Measurements quoted in \kms.}
\label{t_RV}
\begin{tabular}{lrrrrrr|rrrrr}
\hline\hline
 & \multicolumn{6}{c}{Hot Components} & \multicolumn{5}{c}{Cool Components} \\
\hline 
MJD & \ion{He}{i} & \ion{He}{ii} & \ion{Si}{iii} & \ion{Si}{iv} & \ion{N}{ii} & \ion{N}{iii} & \ion{He}{i} & \ion{Si}{ii} & \ion{Si}{iii} & \ion{N}{ii} & \ion{Mg}{ii} \\
\hline
54804 & $-$84$\pm$14 & - & - & $-$86$\pm$11 & $-$99$\pm$20 & $-$91$\pm$4 & $-$7$\pm$10 & 4$\pm$11 & - & $-$40$\pm$20 & $-$14$\pm$10 \\
54808 & - & 267$\pm$10 & 279$\pm$10 & - & 287$\pm$10 & - & - & - & 54$\pm$10 & $-$4$\pm\phantom{1}$3& -  \\
54836 & - & - & - & $-$12$\pm$11 & - & $-$22$\pm$17 & $$0$\pm$15 & - & - & 15$\pm$10 & -  \\
54867 & $-$106$\pm$24 & -  & - & $-$129$\pm$10 & $-$145$\pm$20 & $-$110$\pm$26 & $-$20$\pm$12 & $-$5$\pm\phantom{1}$9 & - & $-$67$\pm$20 & $-$14$\pm$13  \\
55108 & $-$96$\pm$29 & -  & - & $-$142$\pm$11 & $-$178$\pm$20 & $-$136$\pm$16 & $-$13$\pm$13 & $-$15$\pm$12 & - & $-$66$\pm$20 & $-$29$\pm$10 \\
55114 & 103$\pm$21 & -  & - & 130$\pm$10 & 56$\pm$20 & 100$\pm$13 & 17$\pm$11 & 18$\pm$11  & - & 7$\pm$20 & 47$\pm$10 \\
\hline
\end{tabular}
\end{center}
\end{table*}

\subsection{Projected rotational velocities} 	\label{s_vsini}

Assuming that the hot and cool components originate in stellar
photospheres, we can attempt to estimate their projected rotational
velocities.  We have used a Fourier transform (FT) approach as discussed
by \citet{sim07}, supplemented by fitting rotationally-broadened
theoretical line profiles (PF) to the observed spectral features. For the latter we
assumed that rotation dominated other broadening mechanisms and hence
these are best considered as upper limits. Further details on the profile-fitting methodology can be
found in \citet{rya02}.

For the hot component we used the \ion{Si}{iv} lines at 4089 and
4116\AA\ and the \ion{N}{iii} line at 4097\AA. For those LR02 epochs
where the lines were discernible from features in the cool or circumstellar
spectra, the data were shifted to a common rest frame and combined.
This led to relatively clean profiles for the \ion{Si}{iv} lines but
the \ion{N}{iii} line was still affected on its red wing by H$\delta$
emission. The estimates are listed in Table \ref{t_vsini} and are in
reasonable agreement for both methodologies. Additionally, as expected,
the PF estimates (which are upper limits) are generally larger than
those from the FT approach. This provides indirect support for the
validity of the latter. A simple average of the FT results leads
to a value of 66\,\kms for the hot component, with the spread of estimates and the upper
limit from the PF estimates implying a conservative error bar of
$\pm$10\,\kms.

For the cool component, lines due to several metal species were
identified. These suffered from blending with either the hot or
circumstellar components. This was exacerbated by the features having
a small RV variation and therefore not shifting with respect to the
circumstellar features between epochs. Hence the stronger \ion{He}{i}
features were used, although their analysis is complicated by their
greater intrinsic broadening. Additionally, several lines showed
P~Cygni profiles (see Sect. \ref{sp_B}) and were not considered. The
estimates are summarised in Table \ref{t_vsini}. They exhibit the
expected systematic shift between the PF and FT approaches and lead to
a mean FT estimate of 53$\pm$10\,\kms for the cool component; the error estimate again arises
from the spread of values and the upper limit implied by the PF
results.

\begin{table}
\caption{Estimates of the projected rotational velocity ($v$sin$i$) of the cool and hot components from the 
profile fitting (PF) and Fourier transform (FT) methods.}
\label{t_vsini}
\begin{center}
\begin{tabular}{lccc}
\hline\hline
Component & Feature & \multicolumn{2}{c}{$v$sin$i$} \\
& & PF &FT  \\
& & (\kms) & (\kms) \\
\hline
Hot    & \ion{Si}{iv} 4089\AA & 79 & 63\\
       & \ion{Si}{iv} 4116\AA & 68 & 61\\
       & \ion{N}{iii} 4097\AA & 77 & 74\\
\\
Cool  & \ion{He}{i} 4009\AA & 70& 60\\
       & \ion{He}{i} 4143\AA & 64 & 50\\
       & \ion{He}{i} 4437\AA & 58 & 49\\
\hline
\end{tabular}
\end{center}
\end{table}

\section{Atmospheric parameters}	  	\label{s_stellar}

The complexity of the VFTS698 spectrum, containing several different components, makes any quantitative analysis both difficult and prone to systematic error. However the resulting quantitative information could be useful despite the large uncertainties. We have therefore attempted to obtain constraints on the effective temperature, surface gravity and nitrogen abundances for the two components using the non-LTE model atmosphere grid {\sc tlusty} \citep{hub88}; further details of the grids can be found in \cite{rya03} and \citet{duf05}. 

\subsection{Hot component}

The well-resolved spectral features of the hot stellar component made estimating its effective temperature from the \ion{Si}{iii}/\ion{Si}{iv} ionisation equilibrium relatively straightforward.  Due to the strong \ion{Fe}{ii} emission and absorption spectrum between 4500 and 4560\AA\,(shown in Fig. \ref{f_698_R4_R50}), as well as the wavelength cut-off at 4560\AA, the \ion{Si}{iii} lines were not measurable in the LR02 setting.  The longer wavelength coverage of the LR03 setting and the radial velocity of the hot component ($>$200\,\kms) in these data allowed equivalent-width measurements of \ion{Si}{iii} 4567\AA\,for features that had a RV consistent with the hot and cool photospheric spectra.  The equivalent widths of \ion{Si}{iv} were taken from measurements of the 4089 and 4116\AA\,lines, in the LR02 data.  The two \ion{Si}{iv} lines displayed variations in intensity between epochs of 12\% and 26\%, respectively.  The average equivalent widths for both \ion{Si}{iv} lines gave an effective temperature estimate of $\sim$27\,000\,K.  However, the equivalent widths will be underestimated due to contributions from the other components in the spectrum.  We assumed that this contribution was the same for all lines, which is not unreasonable given the relatively small wavelength separation. Then we investigated its effect on the effective temperature estimate by scaling the equivalent widths.  Doubling the equivalent widths led to an increase in the estimate of 1\,000\,K.

A lower limit to the effective temperature of the hot component was estimated from the \ion{He}{ii} absorption line at 4686\AA.  Rotationally-broadened profiles with a \vsini of 70\,\kms were generated from the {\sc tlusty} grid of model atmospheres.  From a $\chi^2$ fit an effective temperature of $\sim$ 24\,000\,K was found together with a gravity, \logg $\sim$ 3.3.  If the hot spectra contributed 50\% of the continuum flux this estimate would be increased by 1\,500\,K. The uncertainty in these estimates is difficult to judge. A lower limit of 24\,000K would appear reasonable both from the \ion{He}{ii} spectrum and the presence of strong \ion{Si}{iv} lines. An upper limit of 28\,000\,K is reasonable from the well observed \ion{Si}{iii}. Hence we adopt 26\,000 $\pm$ 2\,000\,K, but acknowledge that there remains the possibility of additional systematic errors from our assumption of a classical non-LTE photospheric model.


\subsection{Cool component}

For the cool photospheric spectrum, the detection of \ion{Si}{ii} and
\ion{Si}{iii} lines also allows the silicon ionisation balance to be
used to estimate the effective temperature.  Equivalent-width
estimates were obtained for the \ion{Si}{ii} 4128 and 4130\AA\/
absorption lines from the LR02 data, and for
\ion{Si}{iii} at 4567\AA.  Those for the former were difficult to
estimate as the RVs were inconsistent within an epoch
(see Table \ref{t_RV}).  This discrepancy is probably due to a weak emission feature around 4129\AA,
and a \ion{Cr}{ii} absorption component seen to be blended with the
\ion{Si}{ii} 4130\AA\/ line.  Assuming there is no flux contamination for
the other spectral components, this leads to an effective temperature
estimate of 17\,800\,K, for an assumed logarithmic gravity of \logg = 3.0
based on other B-type supergiants taken from \citet{hun08a}.  Making
similar assumptions as in the analysis of the hot component, i.e.
doubling the equivalent widths, leads to an estimate of
18\,400\,K. Again the uncertainty is difficult to estimate. However, 
the clear presence of both ionisation states would imply that, 
excluding unknown systematic errors, an error of $\pm$2\,000\,K would
appear to be appropriate and therefore we adopt a \teff of 18\,000 $\pm$
2\,000\,K. 

Apart from a weak constraint from the profile fitting of the
\ion{He}{ii} line, it was not possible to estimate the surface gravity
of either components due to the strong emission in the Balmer lines.

In Sect. \ref{s_classification} a limit was placed on the luminosity
of VFTS698 assuming it to be a single source.  However, informed by the
estimates of effective temperature further constraints could be placed
on each of the components.  Both the hot and cool spectra appear to be
prominent, hence their contribution to the overall spectrum is thought
to be comparable.  Based on the effective temperature estimates from
Table \ref{t_summary} we deduce luminosities for the hot and cool
components of log~$L/L_{\odot}$ = 5.3 and 5.0, respectively, using
bolometric corrections from \citep{cro06} and identical absolute
visual magnitudes.  To consider these components as main sequence objects would lead to a mass ratio estimate of $\sim$2, but given our analysis of the VFTS698 system classifies it as a sgB[e] system this consideration is not applicable.


\subsection{Nitrogen abundance estimates} \label{s_nitrogen}

The photospheric abundance of nitrogen is important for constraining the amount of nucleosynthetically-processed material present in a stellar atmosphere \citep{heg00, prz10, bro11a}.  As \ion{N}{ii} lines from both stellar components could be identified in the spectra, we can set lower limits on their nitrogen abundances.

For the hot component, adopting an effective temperature of
26\,000\,K, a gravity of \logg = 3.3 and a microturbulence of 10\,\kms
leads to an estimate of $\sim$7.3\,dex from three \ion{N}{ii} lines at
4600\AA.  Note that adopting quantities for the gravity and
microturbulence that would be expected for a star on the main sequence
results in an effective temperature estimate of 30\,000\,K and a
surface nitrogen abundance of $\sim$8.0\,dex.  These weaker
transitions were initially chosen because of their large velocity
separation. The stronger \ion{N}{ii} line at 3995\AA\, was identified
in all the LR02 epochs, but the blending of the two stellar signatures
made it difficult to disentangle the equivalent width estimates.
Following a similar procedure to that used in measuring the
RVs of the \ion{He}{i} profiles, two-component Gaussian fits were
employed, with the equivalent width of the hot component taken to be that with a
RV component corresponding to the higher-excitation metal
lines.  Although there was a variation of the line strength between
epochs, a mean value of 50\,m\AA\,\,implied a nitrogen abundance of
7.2\,dex using the same atmospheric parameters.

The above analysis assumes that all the continuum flux arises from the hot photospheric spectrum and therefore abundances were also estimated assuming 50 and 25\% of the flux came from the hot component, as summarised in Table \ref{t_nitrogen}.  The fact that both spectra are clearly visible would suggest each component might contribute $\sim$50\% to the total flux, which would then suggest a nitrogen abundance of 8.0\,dex for the hot component.  However, regardless of the flux contribution, there would appear to be an enrichment of nitrogen as, at the limit of a 100\% contribution, the estimates show a 0.6\,dex enhancement.


The same methodology has been used to estimate a nitrogen abundance for the cool spectral features although in some cases the equivalent-width measurements lay outside the theoretical grid at the 25\% flux level.  The equivalent-width estimates of the 3995\AA\, line were found to vary more between epochs than for the hot spectra but, if we adopt the average value, we estimate the lower limit to the nitrogen abundance as 7.9\,dex, i.e. an enrichment of 1\,dex.  Adopting a 50\% flux contribution leads to an estimate of $\sim$8.6\,dex.

\begin{table*}
\caption{Equivalent width (EW) measurements (quoted in m\AA) and abundance estimates for both \ion{N}{ii} components.}
\label{t_nitrogen}
\begin{center}
\begin{tabular}{lccccccccccc}
\hline\hline
Epoch & Flux & \multicolumn{2}{c}{4601} & \multicolumn{2}{c}{4607} & \multicolumn{2}{c}{4613} & \multicolumn{2}{c}{4643}  & \multicolumn{2}{c}{3995}\\
(MJD) & \% & EW & N & EW & N & EW & N & EW & N  & EW & N \\
\hline
\multicolumn{4}{l}{\it Hot component:} \\
\hline
54808 & 100 & $\phantom{1}$34 & 7.6 & $\phantom{1}$23 & 7.5 & ...  & ... & $\phantom{1}$51 & 7.8 & $\phantom{1}$50 & 7.2 \\
  	   & $\phantom{1}$50 & $\phantom{1}$68  & 8.0 & $\phantom{1}$46 & 7.9 & ... & ... & 102 & 8.3 & 100 & 7.6\\
  	   & $\phantom{1}$25 & 136 & 8.6 & $\phantom{1}$92 & 8.4 & ... & ... & 204 & \tablefootmark{*} & 200 & 8.3\\
\hline
\multicolumn{4}{l}{\it Cool component:} \\
\hline
54808 & 100 & 110 & 8.3 & 120 & 8.5 & $\phantom{1}$68  & 8.2 & ... & ... & 170 & 7.9 \\
	   & $\phantom{1}$75  & 147 & 8.5 & 160 & 8.6 & $\phantom{1}$91 & 8.4 & ... & ... & 230 & 8.2 \\
  	   & $\phantom{1}$50 & 220 & 8.6 & 240 & 8.5 & 136 & 8.6 & ... & ... & 340 & 8.6 \\
\hline
\end{tabular}
\tablefoot{Each EW was measured assuming different amounts of photospheric flux.  For both components the adopted atmospheric parameters were taken from Table \ref{t_summary}.
\tablefoottext{*}{denotes the EW is outside of the theoretical grid.}
}
\end{center}
\end{table*}

Adopting a 50\% flux contribution from the hot and cool components leads to silicon abundances between 7.0 and 7.2 dex which are in good agreement with those found for early-type stars in the LMC \citep[see, e.g.,][]{hun08a}, providing some indirect support for the methodology adopted. However we stress that the estimates of the atmospheric parameters and nitrogen abundances should be treated with caution because of the complexity of the VFTS698 spectra.  In order to discuss the evolutionary status of VFTS698 in the following sections, we provide a summary table of the estimated properties of both hot and cool spectra in Table \ref{t_summary}.

\begin{table*}
\begin{center}
\caption{Summary of derived parameters for the VFTS698 system, assuming 50\% of the flux comes from each component.}
\label{t_summary}
\begin{tabular}{lccccccccc}
\hline\hline
Component & RV amplitude & \vsini & log~$L/L_{\odot}$ & \teff & \logg & $\xi$ & N & Si  \\
 & (\kms) & (\kms) & & (K) & log~(cms$^{-2}$) & (\kms) & &  \\
\hline
Hot & 210 & 66 & 5.3 & 26\,000 & 3.3 & 10\tablefootmark{*} & 8.0 & 7.0  \\
Cool & $\phantom{1}$16 & 53 & 5.0 & 18\,000 & $\phantom{\,\ast}$3.0\tablefootmark{*} & 20\tablefootmark{*} & 8.6 & 7.2  \\
\hline
\end{tabular}
\tablefoot{
\tablefoottext{*}{Assumed value of \logg or microturbulance}
}
\end{center}
\end{table*}


\section{Discussion}	\label{s_comparison}

The rich emission-line spectrum of VFTS698, dominated by \ion{H}{i},
\ion{Fe}{ii} and [\ion{Fe}{ii}], is characteristic of stars
demonstrating the B[e] phenomenon.  \citet{lam98} describe this
behaviour as being present at a number of evolutionary stages in the
life-cycle of both high- and low-mass single stars and binaries;
essentially corresponding to the presence of a rich,
density-stratified dusty and gaseous circumstellar environment,
excited by a hot ionising source.  We now discuss the nature of the
circumstellar environment and that of the central `engine' of VFTS698
from the observational properties described in Sect.~\ref{s_analysis}.

\subsection{Circumstellar material} \label{s_cir}

The differing line profiles of both the \ion{Fe}{ii} and
[\ion{Fe}{ii}] transitions (e.g.,  Fig. \ref{f_698_R4_R50}) imply that
they arise in kinematically distinct, gaseous regions of the
circumstellar environment.  Indeed, by analogy to classical Be stars,
the shell profiles of the \ion{Fe}{ii} transitions -- characterised by
deep central absorption troughs -- suggest absorption in a flattened
circumstellar disc or torus seen close to edge on.  This is also
analogous to the P Cygni profiles that arise in a spherically
expanding envelope \citep[e.g.][]{dac92}.  Following from this, the
{\em bona fide} P Cygni profiles observed in e.g. the \ion{He}{i}
4026{\AA} transition suggest a spherical outflow from the system.
Such hybrid behaviour is common amongst sgB[e] stars, where it is
often interpreted as resulting from a `composite' wind consisting of a
dense, outflowing circumstellar disc and a high-velocity polar wind
\citep[e.g.][]{zic85}.

However, as demonstrated clearly by Figs.~\ref{f_IR_near} and \ref{f_IR}, the IR properties of VFTS698 differ from those of {\em
  bona fide} sgB[e] stars.  \citet{kas06,kas10} showed that the IR emission from these stars is consistent with a population of warm
grains located within a circumstellar disc with an inner gap resulting in a deficiency of hot dust. Upon consideration of the dust chemistry
they further speculated that these are long-lived quasi-Keplerian discs, with material deposited from a previous red supergiant
phase.  By analogy to similar structures associated with post-AGB binary stars, it is possible that these discs are in fact circumbinary with an (unseen) companion responsible for the disc truncation. In contrast, VFTS698 does not appear to support large quantities of cool dust, although the presence of an apparent near-IR continuum excess raises the possibility of the presence of hot dust; detailed
modelling will be required to characterise the nature of this excess, given that we would also expect a significant free-free emission from
the gaseous component of the circumstellar environment. 

\subsection{Photospheric sources} \label{s_dis_stellar}

\begin{figure}
\begin{center}
\includegraphics[bb = 0 0 581 757, angle=90,scale=.34]{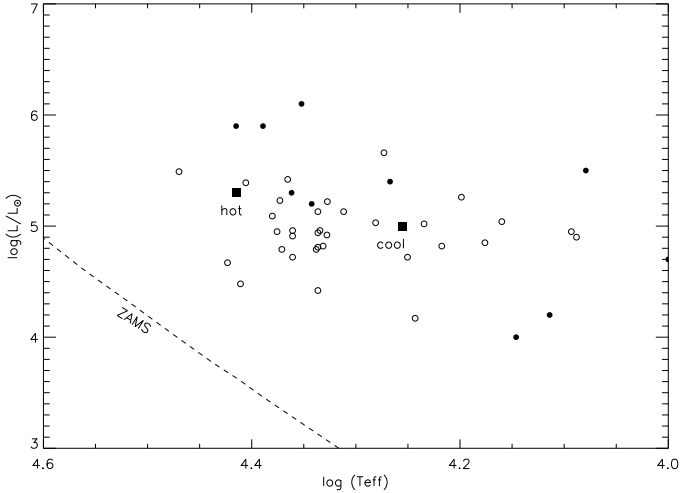}
  \caption{Hertzsprung--Russell diagram comparing the two components of
    VFTS698 (solid squares) with B-type supergiants (open circles)
    from \citet{hun08a} and sgB[e] stars (solid circles) from
    \citet{lam98}.}
\label{f_Be_LMC}
\end{center}
\end{figure}

We next turn to the central, exciting star(s) of VFTS698.
\citet{lam98} summarised the stellar properties of a representative
sample of Magellanic Cloud sgB[e] stars, from investigations by
\citet{zic86} and \citet{gum95}.  The temperatures of both our
putative hot and cool components (Table \ref{t_summary}) are
consistent with these values, although the luminosities we infer
are, for the most part, towards the lower range demonstrated (cf.
Sect.~\ref{s_stellar}), with 5.3 and 5.0\,dex for the hot and cool
components, respectively.  The luminosity and effective temperature
estimates of VFTS698 and other B[e] and B-type supergiants are illustrated
on the Hertzsprung--Russell (H--R) diagram shown in Fig.~\ref{f_Be_LMC}.  For
the majority of B[e] objects it is evident that both components of VFTS698
appear at the lower luminosity end, but are brighter than the majority of
B-type supergiants.  The evolutionary nature of these lower luminosity
(and mass) B[e] stars is uncertain \citep[e.g.][]{gum95} but we note
that this phenomenon is not uncommon or confined solely to the
Magellanic Clouds, with a population also identified within M33 \citep{cla12}.

A key observational constraint placed on the stellar component(s) of
VFTS698 is the presence of pronounced spectroscopic and photometric
variability.  The presence of variability of the \ion{Si}{iv} lines
echoes the findings of \citet{zic86}, who detected \ion{Si}{iv} 4089
and 4116\AA\/ absorption in the spectrum of Hen S22 where previously
\citet{mur78} observed no such features (although the source of this
variation remains unclear).  More telling is the presence of RV shifts
in lines associated with both hot and cool components. Such behaviour
is characteristic of both stellar pulsations and binary motion,
although the magnitude of the RV changes in the hot component is
significantly greater than expected from pulsations alone
\citep{cla10,cla12}.  Indeed, with two plasmas with significantly
different temperatures (supported by the wide range of ionisation
states observed, e.g., \ion{Si}{ii}/\ion{Si}{iii}/\ion{Si}{iv} and
\ion{N}{ii}/\ion{N}{iii}), with very distinct RV behaviours, it is
highly plausible that VFTS698 is a hierarchical stellar system.

The multi-year photometric dataset strongly supports such an
assertion, revealing a persistent 12.7\,d periodicity superimposed
over a long-term variation for a duration of 8 years (Sect.
\ref{s_photo_analysis}).  While pulsationally-induced photometric
variability appears ubiquitous across the upper reaches of the H--R
diagram \citep[e.g.][]{cla12} the pulsations observed in the so-called
supergiant $\alpha$ Cygni variables do not replicate the {\em strict
  periodicity} observed in VFTS698 over the 8\,year observational
period.  Specifically, we may contrast this behaviour to that of the
Galactic B[e] star GG Car, which \citet{gos84} showed to have a
photometric period of 31.02\,d, with an amplitude of $\sim$0.5\,mag
that is stable over a 4\,year period.  Subsequent spectroscopic
observations \citep{gos85} revealed correlated RV changes suggesting a
binary nature for this system.  Unfortunately, with only six epochs of
LR02 spectroscopy for VFTS698, we are unable to determine accurate
periodicities, although, for completeness, we note that the RV
variations were found in Sect.~\ref{s_lines} to be reasonably
compatible with the 12.7\,d photometric period.


\subsection{Interacting binary} \label{s_binary}

Under the assumption of binarity, what may we conclude about the
properties of the VFTS698 system? Its photometric properties suggest
obvious similarities with the class of DPVs identified by
\citet{men03} from OGLE-II data.  These intrinsically blue objects
demonstrate both short- ($P_{1}$, days) and long-term periodicities
($P_{2}$, hundreds of days).  \citeauthor{men03} suggested that these
are related to the orbital period and precession of an elliptical
accretion disc and/or episodic mass loss, respectively, within a
semi-detached binary system\footnote{\citeauthor{men03} relate the
  periodicities in DPVs by $P_{1}$\,$=$\,35.2\,$\pm$\,0.8\,$P_{2}$;
  later revised by \citet{men09b} to $\sim$33.  From our period search
  of VFTS698 (Sect. \ref{s_photo_analysis}) we find a 12.7\,d
  period, and some evidence for a 400\,d period, leading to a ratio
  of $\sim$31.}.  As such they would be closely related in
evolutionary terms to the better documented W~Serpentis/$\beta$ Lyrae
binaries in the Galaxy \citep[e.g.][and references therein]{tar00}.
In these short-period\footnote{E.g.: $\beta$ Lyrae (12.9\,d),
  W~Serpentis (14.16\,d), V367~Cyg (18.16\,d), from
  \citet{tar00}.}, interacting systems the mass loss from the
initially more massive component results in the formation of an
accretion disc/torus (which veils the mass gainer) and in the reversal of
the mass ratio of the system.  A consequence of this configuration is the
presence of photometric modulation on the time-scale of the orbital
period, as well as binary RV variability in the spectrum of the
primary.  Binary motion is also visible in the spectroscopic signature
of the accretion disc around the mass gainer {\em in the subset of
  systems where it may be isolated} \citep[cf.][]{tar00}, while some
systems also demonstrate long term, aperiodic, photometric variability
\citep[e.g.][]{str87}.  Finally, the mass transfer drives a complex
circumstellar environment, with evidence for both polar mass-loss
\citep[e.g. $\beta$ Lyrae; ][]{har96} and the presence of a
circumbinary disc \citep[e.g. BY Cru; ][]{tar00} in some systems.
While the presence of P Cygni profiles in selected \ion{He}{i}
transitions of VFTS698 is suggestive of the former, the lack of RV
variability in the \ion{Fe}{ii}/[\ion{Fe}{ii}] emission-line spectrum
argues for dynamical decoupling of the material from which it
originates.

Since VFTS698 appears to demonstrate the majority of the physical
phenomena associated with W~Serpentis-type, semi-detached eclipsing
binaries, we consider such an identification as highly promising.  A
particularly interesting comparator is the massive
($\sim$7\Msun+30\Msun) binary RY Scuti, due to the similar
orbital period ($\sim$11\,d modulation present in both photometry and
spectroscopy) and spectral type of the mass donor/secondary
\citep[O9.7 Ibpe;][]{wal82} compared to the hot component in VFTS698
and presence of circumbinary ejecta.  Although the primary/mass gainer
in RY Scuti is enshrouded by circumstellar material, using tomographic
reconstruction \citet{gru07} estimated a B0.5~I classification for the
``massive companion'' but suggest that the latter is actually the
spectrum of the photosphere due to an accretion torus directly
associated with it.  The RV estimates for the hot component of RY
Scuti were found using high-ionisation metal lines (\ion{Si}{iv} and
\ion{N}{iii}), whereas the wings of H$_\alpha$, \ion{He}{ii}
4686\AA\/ emission, and \ion{Si}{iii} absorption were the only
features capable of reproducing the anti-phase for the cool component.
Unfortunately, due to the singular LR03 and HR15N observations for
VFTS698, an equivalent RV analysis is not possible.
Furthermore, as discussed in Sect.  \ref{s_stellar}, \ion{Si}{iii} was
detected for the cool component but because of rich \ion{Fe}{ii}
emission around 4550\AA, multi-epoch data of this ionisation state was
unavailable.  Nevertheless, by analogy to RY Scuti it might be
supposed that the spectroscopic features of the cool companion in
VFTS698 actually arise in the photosphere of an accretion disc/torus
around the veiled mass gainer.  In such a scenario one would
anticipate that RV changes be anti-correlated with those observed in
the hot component, an effect that does not appear to be present in our
limited dataset.  However, we note that if the spectrum of the cool
component was formed at the L1 Lagrange point (where Roche lobe
overflow will occur -- presumably as the material will be at its most
dense there), then this would lie inside the centre of mass of the
system, and so we would expect to see a weak correlation between the
hot and cool component spectra, as is observed.

Further RV observations of VFTS698 are required to investigate this discrepancy. If confirmed, one possible explanation might be the presence of a third body in the system, although in this case one would expect a significantly longer period than the {\em provisional} $\sim$12.7$\pm$0.1\,d period determined from the current dataset.  Moreover, in comparison to RY Scuti, the richer \ion{Fe}{ii}/[\ion{Fe}{ii}] emission-line spectrum suggests differences in the geometry/composition of the circumstellar/binary envelope.

Although no B[e] phenomena has been detected in the comparison interacting binaries\footnote{a possible exception being the SMC star N\,82, which \citet[][]{hey90} determined to be a sgB[e], but this classification was weakend by the investigation of \citet{men10b} from an interacting binary nature.}, it's suggested the SMC sgB[e] R4 is a binary merger, the product of a close binary interaction \citep{zic96,pod06}.  \citet{lan97} have attempted to explain this evolutionary scenario by the following binary interaction.  When the primary evolves and fills its Roche lobe, accreting matter to the secondary, if the accretion time-scale, $\tau_{\dot{M}} = M/\dot{M}$, is shorter than the thermal time-scale of the secondary, $\tau_{KH} = GM^2/(RL)$, both components can fill their Roche lobes, resulting in a contact binary and L2 Roche-lobe overflow.  The material passing through the L2 Lagrangian point, will remove angular momentum from the system, forming a disk-like structure around what will become a merger remnant of an interacting binary.  This would imply that VFTS698 and R4 are in respectively a pre- and post-merger evolutionary state.  However the alternative scenario sees the loss of angular momentum slow the system down enough to prevent a binary merger as the initially more massive companion will explode as a type Ib/c supernova. The result sees the mass donor slim down to a Wolf--Rayet star, leaving a short-period binary system as found by \citet{cla11} in Westerlund 1 \citep[further details can be found in][]{ric12}.

In Sect. \ref{s_nitrogen} the nitrogen abundances were investigated
assuming each component to be that of a stellar photosphere.  However,
in view of the above scenario, it is difficult to ascertain what constraints
can be placed on the analysis of nitrogen abundances.  For the hot
component, our nitrogen abundance estimate of 8.0\,dex still appears
to be a reasonable result given the assumptions made for surface
gravity and microturbulence.  Considering the cool component is
thought to originate from an optically thick disc, our assumption of a
supergiant surface gravity can no longer be applied, and hence it is
unclear what conclusions we can draw on the parameters of the
pseudo-photosphere other than there is evidence for a temperature
consistent with a mid-B type photosphere, with spectral evidence for a
large nitrogen enrichment.

\section{Conclusions} \label{s_conclusions}

A detailed spectroscopic and photometric study of VFTS698 has been presented to ascertain its evolutionary status. In summary, VFTS698, shows evidence for being a double-lined spectroscopic binary,
consisting of early- and mid-B type components.  The spectrum is rich with iron-group emission and absorption features which appear
synonymous with the B[e] phenomenon \citep[e.g.,][]{lam98}.  OGLE-III photometry shows a stable 12.7\,d period
from 8 years of observations, and IRSF and {\it Spitzer} imaging shows VFTS698 to have strong near- and mid-IR excesses.  We present the scenario for a W~Serpentis-type interacting binary based on the following attributes:

\begin{itemize}
\item[1]{The near- and mid-IR excess of VFTS698 appears different to that observed in {\em bona fide} sgB[e] stars;}
\item[2]{The stable 12.7\,d photometric period is consistent with both the orbital motion and short-term orbital periods of W Serpentis-like objects.  RVs of both components appear consistent with this short-period orbit;}
\item[3]{There is evidence for long-term photometric variations, a defining characteristic for this class of binaries \citep{str87};}
\item[4]{A complex circumstellar structure is inferred from the time-variable emission lines which display diverse morphologies, e.g., P~Cygni \ion{He}{i} profiles, and shell-like \ion{Fe}{II} lines, similar to those in $\beta$ Lyrae \citep[see][]{tar00}.}
\end{itemize}

The following points require contemporaneous photometry and spectroscopy, together with detailed modelling in order to better ascertain the exact nature of the VFTS698 system.

\begin{itemize}
\item[1]{The near- and mid-IR excess, iron-group shell spectra, and P Cygni structure in the helium spectra are indicators for a complex circumstellar geometry consisting of circumbinary material, a circumstellar disc and wind outflows.}

\item[2]{The RV measurements of the primary/circumstellar disc appear inconsistent with an anti-correlation which would be expected for a binary system.}
\end{itemize}

Through the course of this investigation, a number of different comparison objects (see Table \ref{t_comparison}) have been used to conclude that VFTS698 is an interacting binary, comprising a visible, hot secondary in orbit with a veiled, cool, massive primary.  We surmise the spectral component associated with the primary to originate from the circumstellar disc, but require more detailed RV measurements and contemporary photometry to support this argument.  

\begin{table*}
\begin{center}
\caption{Comparison of orbital and stellar parameters of VFTS698 derived here with those for known sgB[e] stars and interacting binary systems.}
\label{t_comparison}
\begin{tabular}{lccccccccccc}
\hline\hline
System & Galaxy  & Orb. period & \multicolumn{2}{c}{Spectral type}  & \multicolumn{2}{c}{log~$L/L_{\odot}$} & \multicolumn{2}{c}{\teff} & \multicolumn{2}{c}{\logg} & Ref.\\
 & & (days) & & & & & \multicolumn{2}{c}{(kK)} & \multicolumn{2}{c}{log~(cms$^{-2}$)} & \\
 & & & Pri. & Sec. & Pri. & Sec. & Pri. & Sec. & Pri. & Sec. & \\
\hline
VFTS698 & LMC & 12.7 & mid B & early B & 5.0 & 5.3 & 18.0 & 26.0 & $\phantom{\ast}$3.0\tablefootmark{*} & 3.3 & This work \\
\hline
\multicolumn{11}{l}{\it Comparison B[e] stars:} \\
\hline
Hen S22 & LMC & ... & B0 -- B0.5 & ... & 5.9 & ... & 23.0 & ... & ... & ... & 1 \\
Hen S35 & LMC & ... & B1[e]~Iab & ... & 5.2 & ... & 22.0 & ... & 3.0 & ... & 2 \\
R4 & SMC &  7774.5 & B0.5 & A & 5.0 & 4.2 & 27.0 & 9.5 & 3.2 & 2.5 & 3 \\
R50 & SMC & ... & B2 -- B3 & ... & 5.7 & ... & 17.0 & ... & ... & ... & 1 \\
\hline
\multicolumn{11}{l}{\it Comparison interacting binaries:} \\ 
\hline
RY Scuti & MW & 11.1 & O9.7~Ibpe & B0.5~I & $>$5.0$\phantom{>}$ & $>$5.0$\phantom{>}$ & 27.0 -- 30.0 & 27.0 -- 30.0 & 3.0 & 3.0 & 4,5 \\
$\beta$ Lyrae & MW & 12.9 & B? & B6-8~II & 4.4 & 3.8 & 30.0 & 15.0 & ... & ... & 6,7 \\
\hline
\end{tabular}
\tablefoot{
\tablefoottext{*}{assumed value}
\tablefoottext{1}{\citet{zic86}}
\tablefoottext{2}{\citet{gum95}}
\tablefoottext{3}{\citet{zic96}}
\tablefoottext{4}{\citet{gru07}}
\tablefoottext{5}{\citet{men05}}
\tablefoottext{6}{\citet{deg94}}
\tablefoottext{7}{\citet{har96}}
}
\end{center}
\end{table*}

\begin{acknowledgements}
We would like to thank Alceste Bonanos for discussions regarding the periodicity of the OGLE data.  We thank Vanessa Stroud and the Faulkes Telescope Project for use of the FTS observations.  Thanks to Otmar Stahl for supplying the B[e] comparison spectra.  We are grateful to Selma de Mink, Norbert Langer, Myron Smith, Nolan Walborn and Christopher Watson who have offered their time in detailed discussions on the binary nature of VFTS698.  Thanks to the {\it Spitzer} SAGE and IRSF surveys and the Chandra data archive.  PRD would like to acknowledge financial support from the UK Science and Technology Facilities Council and the Department of Education and Learning in Northern Ireland.

\end{acknowledgements}


\onecolumn 
\newpage
{\tiny
\begin{center}
\begin{longtable}{lccccc}
\caption{Line identifications for the three plasma components of the VFTS698 system.  Equivalent width (EW) and 
full-width half-maximum (FWHM) measurements were found by Gaussian-profile fitting, and are taken from the first 
epoch of observations with the LR02 setting.  Identifications marked by an astrerix have insufficient evidence for an accurate identification.} \\	\label{t_lines} \\
\hline\hline
Wavelength & Species & EW & $\Delta$(EW) & FWHM & $\Delta$(FWHM) \\
(\AA)              &                &     (m\AA)       &     (m\AA)            & (\AA) & (\AA) \\\endfirsthead
\caption{\it{continued}} \\
\hline
Wavelength & Species & EW & $\Delta$(EW) & FWHM & $\Delta$(FWHM) \\
(\AA)              &                &     (m\AA)       &     (m\AA)            & (\AA) & (\AA) \\
\hline
\endhead
\hline
\multicolumn{6}{r}{\it{continued on next page}} \\
\endfoot
\hline
\endlastfoot
\hline
\hline
\multicolumn{6}{l}{\it Lines orginating from a stationary plasma:}\\
\hline
4002.1 & \ion{Fe}{ii} & $\phantom{1}$67.9 & 23.5  & 0.7 & 0.1 \\
4005.2 & \ion{Fe}{ii} & 128.0 & 25.9 & 1.6 & 0.2 \\
4012.4 & \ion{Ti}{ii} & $\phantom{1}$78.1 & 14.6 & 0.7 & 0.1 \\
4015.2 & \ion{Fe}{ii} & $\phantom{1}$60.5 & 16.2 & 0.8 & 0.2 \\
4033.0 & \ion{Fe}{ii} & $\phantom{1}$49.5 & $\phantom{1}$9.0 & 0.7 & 0.1 \\
4042.4 & \ion{Cr}{ii} & $\phantom{1}$43.0 & $\phantom{1}$6.7 & 0.7 & 0.1 \\
4048.8 & \ion{Fe}{ii} & $\phantom{1}$95.2 & $\phantom{1}$7.6 & 0.8 & 0.1 \\
4052.0 & \ion{Cr}{ii} & $\phantom{1}$51.8 & 33.1 & 0.7 & 0.2 \\
4057.5 & \ion{Fe}{ii} & $\phantom{1}$53.2 & $\phantom{1}$6.9 & 0.6 & 0.1 \\
4063.9 & \ion{Cr}{ii} & $\phantom{1}$43.0 & $\phantom{1}$5.7 & 0.6 & 0.1 \\
4077.5 & \ion{Cr}{ii} & $\phantom{1}$24.1 & $\phantom{1}$6.0 & 0.6 & 0.1 \\
4122.6 & \ion{Fe}{ii} & $\phantom{1}$92.0 & $\phantom{1}$8.8 & 0.6 & 0.0 \\
4132.4 & \ion{Cr}{ii} & $\phantom{1}$49.5 & 12.3 & 0.9 & 0.2 \\
4163.6 & \ion{Ti}{ii} & $\phantom{1}$24.8 & $\phantom{1}$8.0 & 0.5 & 0.1 \\
4173.5 & \ion{Fe}{ii} & 136.3 & 13.8 & 0.6 & 0.0 \\
4178.9 & \ion{Fe}{ii} & 103.7 & 13.0 & 0.5 & 0.1 \\
4195.4 & \ion{Cr}{ii} & $\phantom{1}$11.7 & $\phantom{1}$5.2 & 0.6 & 0.2 \\
4244.8 & [\ion{Fe}{ii}] & ... & ... & ... & ... \\
4261.8 & \ion{Cr}{ii} & $\phantom{1}$95.0 & $\phantom{1}$6.4 & 0.7 & 0.0 \\
4269.3 & \ion{Cr}{ii} & $\phantom{1}$45.6 & $\phantom{1}$6.4 & 0.7 & 0.1 \\
4271.9 & \ion{Ti}{ii} & $\phantom{1}$55.0 & $\phantom{1}$6.8 & 0.8 & 0.1 \\
4273.3 & \ion{Fe}{ii} & 100.0 & $\phantom{1}$6.2 & 0.6 & 0.0 \\
4274.1 & \ion{Ti}{ii} & 100.0 & $\phantom{1}$6.2 & 0.6 & 0.0 \\
4275.6 & \ion{Cr}{ii} & 159.6 & $\phantom{1}$7.9 & 1.0 & 0.0 \\
4276.8 & [\ion{Fe}{ii}] & ... & ... & ... & ... \\
4278.1 & \ion{Fe}{ii} & $\phantom{1}$87.1 & $\phantom{1}$6.9 & 0.8 & 0.0 \\
4287.4 & [\ion{Fe}{ii}] & ... & ... & ... & ... \\
4290.2 & \ion{Ti}{ii} & $\phantom{1}$78.9 & $\phantom{1}$9.5 & 0.8 & 0.1 \\
4296.6 & \ion{Fe}{ii} & 146.7 & 12.4 & 0.6 & 0.0 \\
4300.1 & \ion{Ti}{ii} & $\phantom{1}$57.0 & 27.2 & 1.1 & 0.5 \\
4301.9 & \ion{Ti}{ii} & $\phantom{1}$26.2 & $\phantom{1}$9.3 & 0.6 & 0.2 \\
4303.2 & \ion{Fe}{ii} & 162.8 & 12.3 & 0.6 & 0.0 \\
4308.9 & \ion{Ti}{ii} & $\phantom{1}$82.6 & 12.9 & 0.7 & 0.1 \\
4312.9 & \ion{Ti}{ii} & $\phantom{1}$48.8 & $\phantom{1}$7.9 & 0.7 & 0.1 \\
4314.3 & \ion{Ti}{ii} & $\phantom{1}$37.2 & 11.3 & 0.5 & 0.1 \\
4319.6 & [\ion{Fe}{ii}] & ... & ... & ... & ... \\
4330.7 & \ion{Ti}{ii} & $\phantom{1}$23.5 & $\phantom{1}$4.1 & 0.7 & 0.1 \\
4351.8 & \ion{Fe}{ii} & 188.1 & $\phantom{1}$8.6 & 0.7 & 0.0 \\
4358.1 & [\ion{Fe}{ii}] & ... & ... & ... & ... \\
4369.4 & \ion{Fe}{ii} & $\phantom{1}$40.3 & 10.2 & 0.5 & 0.1 \\
4382.8 & [\ion{Fe}{ii}] & ... & ... & ... & ... \\
4385.4 & \ion{Fe}{ii} & $\phantom{1}$66.7 & 16.0 & 0.4 & 0.1 \\
4414.5 & [\ion{Fe}{ii}] & ... & ... & ... & ... \\
4416.3 & [\ion{Fe}{ii}] & ... & ... & ... & ... \\
4443.8 & \ion{Ti}{ii} &$ \phantom{1}$57.9 & 12.7 & 0.6 & 0.1 \\
4455.9 & \ion{Fe}{ii} & 183.5 & 21.3 & 1.2 & 0.1 \\
4470.3 & [\ion{Fe}{ii}] & ... & ... & ... & ... \\
4488.3 & \ion{Ti}{ii} & 108.8 & 11.2 & 0.6 & 0.0 \\
4488.8 & [\ion{Fe}{ii}] & ... & ... & ... & ... \\
4491.4 & \ion{Fe}{ii} & 153.8 & 11.5 & 0.6 & 0.0 \\
4492.6 & [\ion{Fe}{ii}] & ... & ... & ... & ... \\
4508.3 & \ion{Fe}{ii} & 189.9 & $\phantom{1}$8.9 & 0.7 & 0.0 \\
4509.6 & [\ion{Fe}{ii}] & ... & ... & ... & ... \\
4514.9 & [\ion{Fe}{ii}] & ... & ... & ... & ... \\
4515.3 & \ion{Fe}{ii} & 186.3 & $\phantom{1}$8.9 & 0.7 & 0.0 \\
4520.2 & \ion{Fe}{ii} & 159.3 & 13.2 & 0.6 & 0.0 \\
4522.6 & \ion{Fe}{ii} & 221.9 & 14.3 & 0.7 & 0.0 \\
4534.2 & \ion{Fe}{ii} & 161.6 & 14.4 & 0.7 & 0.0 \\
4541.5 & \ion{Fe}{ii} & 149.9 & 14.3 & 0.7 & 0.1 \\
4549.5 & \ion{Fe}{ii} & 272.4 & 13.9 & 0.7 & 0.0 \\
4550.5 & [\ion{Fe}{ii}] & ... & ... & ... & ... \\
4552.0 & [\ion{Fe}{ii}] & ... & ... & ... & ... \\
4555.0 & [\ion{Fe}{ii}] & ... & ... & ... & ... \\
4555.9 & \ion{Fe}{ii} & $\phantom{1}$64.1 & $\phantom{1}$9.6 & 0.4 & 0.1 \\
4558.7 & \ion{Cr}{ii} & 172.3 & 13.5 & 0.8 & 0.0 \\
4576.3 & \ion{Fe}{ii} & 116.8 & 17.5 & 0.5 & 0.1 \\
4582.8 & \ion{Fe}{ii} & $\phantom{1}$74.8 & 18.3 & 0.8 & 0.1 \\
4583.8 & \ion{Fe}{ii} & 136.2 & 15.4 & 0.5 & 0.1 \\
4588.2 & \ion{Cr}{ii} & 148.8 & 18.3 & 0.7 & 0.1 \\
4592.0 & \ion{Cr}{ii} & $\phantom{1}$81.3 & 16.9 & 0.6 & 0.1 \\
4616.6 & \ion{Cr}{ii} & $\phantom{1}$97.9 & $\phantom{1}$4.3 & 0.7 & 0.1 \\
4618.1 & \ion{Cr}{ii} & 135.2 & $\phantom{1}$4.2 & 0.7 & 0.1 \\
4620.3 & \ion{Cr}{ii} & 114.9 & $\phantom{1}$4.2 & 0.7 & 0.1  \\
4629.3 & \ion{Fe}{ii} & 155.4 & $\phantom{1}$5.5 & 0.6 & 0.1 \\
4634.3 & \ion{Fe}{ii} & 151.3 & $\phantom{1}$6.7 & 0.8 & 0.1 \\
4635.3 & \ion{Fe}{ii} & 126.1 & $\phantom{1}$6.0 & 0.6 & 0.1 \\
4656.8 & \ion{Fe}{ii} & 158.1 & $\phantom{1}$4.9 & 0.8 & 0.1 \\
4658.1 & [\ion{Fe}{iii}] & ... & ... & ... & ... \\
4665.7 & [\ion{Fe}{ii}] & ... & ... & ... & ... \\
4670.3 & \ion{Fe}{ii} & $\phantom{1}$51.1 & $\phantom{1}$4.3 & 0.6 & 0.1 \\
4701.5 & [\ion{Fe}{iii}] & ... & ... & ... & ... \\
4728.1 & [\ion{Fe}{ii}] & ... & ... & ... & ... \\
4733.9 & [\ion{Fe}{iii}] & ... & ... & ... & ... \\
4774.7 & [\ion{Fe}{ii}] & ... & ... & ... & ... \\
4798.3 & [\ion{Fe}{ii}] & ... & ... & ... & ... \\
4805.1 & \ion{Ti}{ii} & $\phantom{1}$23.6 & $\phantom{1}$4.8 & 0.6 & 0.1 \\
4812.3 & \ion{Cr}{ii} & $\phantom{1}$37.9 & $\phantom{1}$3.6 & 0.6 & 0.1 \\
4813.9 & [\ion{Fe}{iii}] & ... & ... & ... & ... \\
4814.5 & [\ion{Fe}{ii}] & ... & ... & ... & ... \\
4824.1 & \ion{Cr}{ii} & 130.3 & $\phantom{1}$4.7 & 0.6 & 0.1 \\
4836.2 & \ion{Cr}{ii} & $\phantom{1}$63.2 & $\phantom{1}$5.0 & 0.7 & 0.1 \\
4848.2 & \ion{Cr}{ii} & 101.5 & $\phantom{1}$8.7 & 0.6 & 0.1 \\
4871.6 & \ion{Fe}{ii} & $\phantom{1}$55.1 & 11.0 & 0.9 & 0.1 \\
4877.5 & \ion{Cr}{ii} & 140.0 & $\phantom{1}$9.5 & 0.7 & 0.1 \\
4889.6 & [\ion{Fe}{ii}] & ... & ... & ... & ... \\
4894.9 & \ion{Cr}{ii} & $\phantom{1}$18.2 & $\phantom{1}$6.7 & 0.5 & 0.1 \\
4901.7 & \ion{Cr}{ii} & $\phantom{1}$46.0 & $\phantom{1}$9.6 & 1.0 & 0.2 \\
4905.4 & [\ion{Fe}{ii}] & ... & ... & ... & ... \\
4926.7 & \ion{Cr}{ii} & ... & ... & ... & ... \\
4952.7 & \ion{Fe}{ii} & $\phantom{1}$28.7 & $\phantom{1}$3.9 & 0.7 & 0.1 \\
4954.0 & \ion{Fe}{ii} & $\phantom{1}$33.4 & $\phantom{1}$3.6 & 0.7 & 0.1 \\
4958.2 & [\ion{Fe}{ii}] & ... & ... & ... & ... \\
5006.7 & [\ion{Fe}{ii}] & ... & ... & ... & ... \\
5019.5 & \ion{Fe}{ii} & 240.9 & $\phantom{1}$9.5 & 0.6 & 0.1 \\
5020.2 & [\ion{Fe}{ii}] & ... & ... & ... & ... \\
6446.4 & \ion{Fe}{ii} & 107.2 & $\phantom{1}$6.4 & 0.6 & 0.1 \\
6456.4 & \ion{Fe}{ii} & 306.8 & $\phantom{1}$6.2 & 0.6 & 0.1 \\
6482.2 & \ion{Fe}{ii} & 106.3 & $\phantom{1}$5.8 & 0.5 & 0.1 \\
6487.3 & \ion{Fe}{ii} & $\phantom{1}$72.3 & $\phantom{1}$6.9 & 0.7 & 0.1 \\
6614.3 & \ion{Fe}{ii} & 127.3 & $\phantom{1}$2.4 & 0.6 & 0.1 \\
6716.2 & \ion{Ti}{ii} & 140.9 & $\phantom{1}$2.5 & 0.6 & 0.1 \\
6731.3 & \ion{Fe}{ii} & $\phantom{1}$95.0 & $\phantom{1}$2.4 & 0.6 & 0.1 \\
6752.0 & \ion{Cr}{ii} & $\phantom{1}$12.5 & $\phantom{1}$2.1 & 0.5 & 0.1 \\
6764.1 & \ion{Fe}{ii} & $\phantom{1}$22.9 & $\phantom{1}$2.8 & 0.8 & 0.1 \\
6799.9 & \ion{Fe}{ii} & $\phantom{1}$12.5 & $\phantom{1}$2.5 & 0.6 & 0.1 \\
\hline
\multicolumn{6}{l}{\it Lines associated with the cool component:} \\
\hline
3995.00 & \ion{N}{ii} & 198.7 & 15.5 & 1.3 & 0.1 \\
4009.26 & \ion{He}{i} & 425.6 & 46.1 & 1.1 & 0.1 \\
4026.19 & \ion{He}{i} & 523.7 & 77.7 & 1.1 & 0.1 \\
4106 & \ion{N}{ii}\tablefootmark{*} & 166.5 & 16.7 & 1.3 & 0.1 \\
4120.82 & \ion{He}{i} & 444.2 & 43.9 & 1.0 & 0.1 \\
4128.05 & \ion{Si}{ii} & 123.9 & 12.8 & 1.2 & 0.1 \\
4130.89 & \ion{Si}{ii} & 122.7 & 10.8 & 0.8 & 0.1 \\
4143.76 & \ion{He}{i} & 421.8 & 50.4 & 1.0 & 0.0 \\
4168.967 & \ion{He}{i} & 187.9 & 12.6 & 1.0 & 0.1 \\
4387.93 & \ion{He}{i} & 555.6 & 24.7 & 1.1 & 0.0 \\
4395.94 & \ion{O}{ii}\tablefootmark{*} & 217.9 & 15.1 & 1.3 & 0.1 \\
4419.6 & \ion{Fe}{iii} & 188.5 & $\phantom{1}$9.9 & 0.9 & 0.0 \\
4431.02 & \ion{Fe}{iii} & 57.6 & 11.4 & 0.7 & 0.1 \\
4437.551 & \ion{He}{i} & 287.3 & $\phantom{1}$4.3 & 1.1 & 0.1 \\
4471.47 & \ion{He}{i} & ... & ... & ... & ... \\
4481.23 & \ion{Mg}{ii} & $\phantom{1}$93.7 & $\phantom{1}$7.5 & 1.1 & 0.1 \\
4552 & \ion{Si}{iii} & 116.9 & 14.5 & 1.4 & 0.1 \\
4567.84 & \ion{Si}{iii} & 157.2 & 48.7 & 2.2 & 0.6 \\
4601.48 & \ion{N}{ii} & 107.7 & 10.6 & 1.8 & 0.1 \\
4607.15 & \ion{N}{ii} & 118.2 & 15.1 & 1.9 & 0.2 \\
4613.87 & \ion{N}{ii} & $\phantom{1}$54.8  & 10.0 & 1.6 & 0.2 \\
4643.09 & \ion{N}{ii} & 275.7 & 15.9 & 2.6 & 0.1 \\
4713.16 & \ion{He}{i} & ... & ... & ... & ... \\
4921.93 & \ion{He}{i} & ... & ... & ... & ... \\
\hline
\multicolumn{6}{l}{\it Lines associated with the hot component:} \\
\hline
4008.00 & \ion{He}{i} & 282.2 & 44.8 & 1.4 & 0.2 \\
4088.84 & \ion{Si}{iv} & 230.8 & 12.6 & 1.6 & 0.1 \\
4097.33 & \ion{N}{iii} & 141.5 & $\phantom{1}$8.7 & 1.3 & 0.1 \\
4116.1 & \ion{Si}{iv} & 135.1 & 12.2 & 1.4 & 0.1 \\
4119 & \ion{He} & 116.5 & 20.3 & 0.8 & 0.1 \\
4143 & \ion{He}{i} & 368.8 & 48.2 & 1.5 & 0.2 \\
4167 & \ion{He}{i} & $\phantom{1}$58.8 & 11.6 & 0.9 & 0.2 \\
4387 & \ion{He}{i} & 164.7 & 21.3 & 0.9 & 0.1 \\
4437 & \ion{He}{i} & $\phantom{1}$91.3 & 20.7 & 1.1 & 0.2 \\
4567.84 & \ion{Si}{iii} & 104.0 & 25.6 & 1.3 & 0.2\\
4601.48 & \ion{N}{ii} & $\phantom{1}$28.8 & $\phantom{1}$9.9 & 1.1 & 0.3 \\
4607.15 & \ion{N}{ii} & $\phantom{1}$24.0 & $\phantom{1}$7.3 & 0.9 & 0.2 \\
4643.09 & \ion{N}{ii} & $\phantom{1}$23.6 & $\phantom{1}$9.3 & 1.0 & 0.3 \\
4685.71 & \ion{He}{ii} & $\phantom{1}$47.3 & $\phantom{1}$7.9 & 2.0 & 0.3 \\
4713.16 & \ion{He}{i} & ... & ... & ... & ... \\
4921.93 & \ion{He}{i} & ... & ... & ... & ... \\
\hline
\end{longtable}
\end{center}
}

\end{document}